\documentclass[12pt,a4paper]{article}
\newcommand{\spacingset}[1]{\renewcommand{\baselinestretch}{#1}\small\normalsize}
\usepackage[english]{babel}
\usepackage[numbers, authoryear, comma, longnamesfirst, sectionbib]{natbib}  
\usepackage{amssymb, amsmath, amsthm, amsfonts}
\usepackage{graphicx, mathtools, booktabs,url}
\usepackage[font=small,labelfont=bf]{caption}
\usepackage[a4paper,hmargin={0.9in,0.9in},vmargin=0.9in]{geometry}
\usepackage{enumitem}

\newenvironment{keywords}
{\par\noindent\textbf{Keywords:}\ }
{\par}

\usepackage{float}
\usepackage{comment}
\usepackage{hyperref}
\usepackage[bottom]{footmisc}
\usepackage{xcolor}
\usepackage{setspace}
\hypersetup{
    colorlinks,
    linkcolor={blue!80!black},
    citecolor={red!80!black},
    urlcolor={blue!80!black}
}

\newtheorem{Theorem}{Theorem}[section]
\newtheorem{Lemma}[Theorem]{Lemma}
\newtheorem{Proposition}[Theorem]{Proposition}
\newtheorem{Definition}[Theorem]{Definition}
\newtheorem{remark}[Theorem]{Remark}

\title{Dynamic Conditional SKEPTIC}
\author{Gabriele Di Luzio\thanks{
Department of Economic and Social Sciences,
Sapienza University of Rome,
 Italy. 
Email: \url{gabriele.diluzio@uniroma1.it}}
\hspace{1cm}
Giacomo Morelli\thanks{
Corresponding author. Department of Statistical Sciences, 
Sapienza University of Rome, 
 Italy.
E-mail: \url{giacomo.morelli@uniroma1.it}}
}

\begin{document}

%\begin{highlights}
%\item  Dynamic Conditional SKEPTIC robustly estimates time-varying correlations.
%\item Rank-based dependence relaxes normality under skewness and heavy tails.
%\item Simulations show more stable parameter estimates for the Dynamic Conditional SKEPTIC than DCC-based models.
%\end{highlights}
\spacingset{1.5}  % per interlinea 1.5
\maketitle
\thispagestyle{empty} % Rimuove intestazione/paginazione
\vspace{-1cm} % Riduce spaziatura 
\begin{abstract}
We introduce the Dynamic Conditional SKEPTIC (DCS), a semiparametric approach for efficiently and robustly estimating time-varying correlations in multivariate models. 
We exploit nonparametric rank-based statistics, namely Spearman's rho and Kendall's tau, to estimate the unknown correlation matrix 
and discuss the stationarity, $\beta$- and $\rho$- mixing  conditions of the model.
We illustrate the methodology by estimating the time-varying conditional correlation matrix of the stocks included in the S\&P100 and S\&P500 during the period from 02/01/2013 to 23/01/2025. The results show that DCS improves diagnostic checks compared to the classical Dynamic Conditional Correlation (DCC) models, providing uncorrelated and normally distributed residuals. A risk management application shows that global minimum variance portfolios estimated using the DCS model exhibit lower turnover than those based on the DCC and DCC-NL models, while also achieving higher Sharpe ratios for portfolios constructed from S\&P 100 constituents.
\end{abstract}

% Use if graphical abstract is present
%\begin{graphicalabstract}
%\includegraphics{}
%\end{graphicalabstract}

% Research highlights

%\nocite{*}

% Keywords
% Each keyword is seperated by \sep

\maketitle

\section{Introduction}

Multivariate GARCH models 
analyze the dependence structure of a vector of asset returns through the dynamic estimation of its covariance matrix. When the number of assets is large, these models face computational issues in the estimation procedure that returns poor performance due to the curse of dimensionality. 
%Following the introduction of ARCH and GARCH for univariate volatility modeling (\citet{engle1982autoregressive} \citet{bollerslev1986generalized}), numerous attempts have been made to extend these frameworks to %study volatility and correlation dynamics in multivariate time series. 
%since the second moment of the multivariate distribution of financial data plays a crucial role in portfolio selection and risk management. 
%The Multivariate GARCH (MGARCH) models have been proposed, with particular interest in those based on correlation rather than covariance.
%The seminal works are the Constant Conditional Correlation model (CCC) of \citet{bollerslev1990modelling}, the Dynamic Conditional Correlation (DCC) model by
%\citet{engle2002dynamic}, and the Varying Correlation (VC) model by \citet{tse2002multivariate}. 
%These models estimate the correlation matrix thought time using an autoregressive process, exploiting the variance-correlation decomposition of the conditional covariance of asset returns. For a comprehensive review of MGARCH models, see \citet{bauwens2006multivariate}.
%Several approaches in modeling multivariate volatility have been proposed. Seminal works in this area are the Constant Conditional Correlation (CCC) model by , the Dynamic Conditional Correlation (DCC) model by
%Its dynamic modeling is performed by Multivariate GARCH (MGARCH) which is the multivariate extension of the ARCH model of \citet{engle1982autoregressive}, and the GARCH model of \citet{bollerslev1986generalized}, a complete survey of MGARCH model is described in \citet{bauwens2006multivariate}. 
To this end, the recent works of \citet{engle2019large} and \citet{de2022large} robustify the Dynamic Conditional Correlation (DCC)  (\citealp{engle2002dynamic}), one of the most commonly used in the class of multivariate GARCH models, for estimating time-varying correlations in financial assets, including a nonlinear shrinkage (\citealp{ledoit2015spectrum}) that improves the estimation of the unconditional correlation matrix and relying on the composite likelihood method  (\citealp{pakel2021fitting}) that provides computationally feasible estimation of the DCC model in large-dimensional settings.

Standard implementations of the DCC model commonly rely on a Gaussian conditional quasi-likelihood, but the returns are not exactly normally distributed and
this assumption does not adequately capture the empirical stylized facts \citep{cont2001empirical} which are commonly observed in asset return distributions \citep{bai2003kurtosis}, particularly the skewness and leptokurtosis.
%Extreme events in asset returns, driven by market shocks, lead to negative skewness and the presence of outliers. This makes the normal distribution unsuitable for modeling financial returns, 
 % [Furthermore, the Pearson correlation coefficient is highly sensitive to such shocks. \textcolor{blue}{Rephrase}]. 
%\textcolor{blue}{and the Pearson correlation coefficient becomes highly sensitive to such events.}
%Robust correlation measures, such as Spearman’s rho or Kendall’s tau, are less influenced by outliers and are preferred to capture nonlinear dependencies.
%\textcolor{red}{Add here....A good motivation that we should integrate is the statistical foundations that justify the need for SKEPTIC! see Liu et al.(2012). We want to avoid a referee commenting: just replace Gaussian innovation in the DCC with Student's t!}
To formalize the concept, let $\boldsymbol{X}_t=(X_{1,t}, \hdots, X_{p,t})^\prime$ represent the $p$-dimensional random vector of asset returns, the main assumption is that $\boldsymbol{X}_t \mid \mathcal{F}_{t-1}\sim \operatorname{N}(0, H_t)$, where $H_t$ is the covariance matrix, and $\mathcal{F}_{t-1}$ is the information up to $t-1$. 
The difference between the normal distribution and the empirical distribution of $\boldsymbol{X}_t$ 
can lead to misleading estimation, resulting in a wrong model fit. 

To strengthen the credibility of our thesis, we collect data on the stock returns that compose the S\&P 500 index, from 02/01/2013 to 23/01/2025, comprising a total of 429 companies. We perform the Kolmogorov-Smirnov test to compare the empirical distribution of returns against both the normal and Student’s \( t \) distributions. For the Student’s \( t \) distribution, we estimate the degrees of freedom by fitting a GARCH(1,1) model to the returns, assuming Student’s \( t \)-distributed innovations. 
\begin{table}[h]
    \centering
    \caption{\textit{Kolmogorov-Smirnov test results for the normal and Student's \( t \) distributions. In the Table, (N) denotes the normal distribution, while (\( t \)) represents the Student's \( t \) distribution. The values in the Table are the number of stocks that reject the null hypothesis at 1\% and 5\% levels.}}
    \label{tests}
    \begin{tabular}{ccc}
        \toprule
        Significance Level & Kolmogorov-Smirnov (N) & Kolmogorov-Smirnov (\( t \)) \\
        \midrule
        1\% & 429 & 429 \\
        5\% & 429 & 429 \\
        \bottomrule
    \end{tabular}
\end{table}
The test results presented in Table \ref{tests} indicate that none of the returns follow either a normal or a Student’s \( t \) distribution, highlighting the need for specifying an alternative, more suitable distribution for modeling asset returns.

We introduce the Dynamic Conditional SKEPTIC (DCS) model that relaxes the normality assumption by allowing the marginal distributions to exhibit skewness and heavy tails while employing a Gaussian copula framework to model dependence. 
In particular, we consider   $\boldsymbol{X}_t \mid \mathcal{F}_{t-1}\sim \operatorname{NPN}(0,H_t,f)$ where $\operatorname{NPN}$ is the nonparanormal distribution  (\citealp{liu2009nonparanormal}) and the set of functions \( \{f_j\}_{j=1}^p \) transform the random vector \( \boldsymbol{X}_t = \left(X_{1,t}, \ldots, X_{p,t}\right)^\prime \),  into the random vector \( f(\boldsymbol{X}_t) = \left(f_{1}\left(X_{t,1}\right), \ldots, f_{p}\left(X_{p,t}\right)\right)^\prime \), which follows a multivariate Gaussian distribution. 
The nonparanormal distribution depends on the transformation functions \( \{f_j\} \) and the covariance matrix \( H_t \), both estimated from the data.

The nonparanormal approach is particularly appealing when the dimensionality is large, thanks to its robustness against data contamination, driven by extreme market events, that deviate the stock returns from the normal distribution. 
Robustness is achieved  applying monotonic transformations to each marginal, which preserve the original ranking of the variables while aligning the marginal distributions with Gaussian distributions.
This allows the nonparanormal model to exploit normal-based dependence structures without requiring normality in the observed data. 

When data are truly normally distributed, Spearman's rho and Kendall's tau statistics may be used to determine the unknown correlation matrix exploiting their relation with the Pearson correlation coefficient (\citealp{kruskal1958ordinal}). This approach is suggested in the  SKEPTIC\footnote{Spearman's rho and Kendall's tau estimates preempt to infer correlation. }
estimator (\citealp{liu2012high}), a nonparametric and robust estimator of the correlation matrix that is less sensitive to outliers and does not strictly rely on the normality assumption as the Pearson correlation coefficient does.

We provide a theoretical background to the model, deriving the stationarity conditions and showing the $\beta$- and $\rho$- mixing conditions of the process.  
We show that our methodology achieves a convergence rate of $O_P\left(\sqrt{\log(Tp)/T}\right)$ for the estimation of the unknown correlation matrix.

In the empirical analysis, we estimate the time-varying correlation matrix of the stock returns of companies that compose the S\&P 100 and S\&P 500 indices and observe that the DCS model leads to a superior fit compared to the DCC and DCC-NL models,
enhancing information criteria and diagnostic checks. Additionally, we provide numerical simulations, and we find that the estimates of process parameters are reliable in the case of the DCS model.  
We also conduct a risk management application, and we construct minimum variance portfolios using the classical Markowitz approach. Notably, the portfolios estimated with the DCS model exhibit low turnover and favorable Sharpe ratios, highlighting both practical benefits and a more robust representation of the underlying dependence structure.

This paper adds to two main research fields: (i)  the modeling of temporal dependence using a copula function  (\citealp{beare2010copulas}, \citealp{patton2012review}, \citealp{creal2013generalized}, \citealp{patton2013copula} and \citealp{fan2014copulas}), and (ii) the nonlinear modeling of multivariate time series with a copula-based function, (\citealp{chen2006estimation}, \citealp{lee2009copula}, \citealp{remillard2012copula}, \citealp{fan2023estimation} and \citealp{cordoni2024consistent}). In particular, \citet{fan2023estimation} consider a semiparametric Gaussian copula process for estimating the transition matrix within a VAR model, whereas \citet{cordoni2024consistent} use it for causal inference in a high-dimensional time series context. 

The outline of the work is the following. In Section \ref{section_2}, we recall the background, whereas the model is presented in Section \ref{section_3}. We provide a theoretical discussion of the model in Section \ref{section_4} and the estimation procedure in Section \ref{section_5}. We conduct a simulation study in Section \ref{section_6}, whereas in Section \ref{section_7}, we propose an empirical analysis based on Markowitz's global minimum variance portfolios. Section \ref{section_8} concludes.
 
\section{Background}\label{section_2}
\subsection{Notation}
 We assume that \( A = \left[A_{ij}\right] \in \mathbb{R}^{p \times p} \) and \( m = \left(m_1, \ldots, m_p\right)^\prime \in \mathbb{R}^p \). 
 %For simplicity, we omit the temporal dependence in the notation.
For \( 1 \leq q < \infty \), the matrix \( \ell_q \)-operator norm is defined as: $\|A\|_q = \sup_{m \neq 0} \frac{\|A m\|_q}{\|m\|_q}$. For \( q = 1 \) and \( q = \infty \), the norms are:
$
\|A\|_1 = \max_{1 \leq j \leq p} \sum_{i=1}^p |A_{ij}|,
$
and $\|A\|_{\infty} = \max_{1 \leq i \leq p} \sum_{j=1}^p |A_{ij}|$.
The matrix \( \ell_2 \)-operator norm, also known as the spectral norm, is defined as the largest singular value of the matrix \( A \). We also define \( \|A\|_{\max} = \max_{ij} |A_{ij}| \), representing the maximum absolute value of any element in the matrix.
Moreover, \( A \geq 0 \) means that all elements of \( A \) are nonnegative, whereas \( A > 0 \) implies that all elements are strictly positive. The operator \( \operatorname{Vech}(A) \) denotes the \( p(p+1)/2:= p^* \)-dimensional column vector formed by stacking the upper triangular elements of \( A \) in a column-wise manner, excluding redundancies. We define \( \operatorname{diag}(A) = \left[a_{ij} \mathbf{1}(i=j)\right]_{1 \leq i,j \leq p} \), and \( \operatorname{Vecd}(A) = \left[a_{ii}\right]_{1 \leq i \leq p} \), the vector consisting of the diagonal elements of \( A \) in \( \mathbb{R}^p \). We indicate $|A|$ the determinant of the matrix A. 
In conclusion, \( \rho(A) \) represents the spectral radius of \( A \), defined as the largest modulus of its eigenvalues. If \( A \) is positive semidefinite, then \( \rho(A) = \|A\|_2 \).
\subsection{Dynamic Conditional Correlation}
Let $\boldsymbol{X}_t$ denote a stochastic process, and $\mathcal{F}_{t-1}$ the information set at time $t-1$, we assume that:
\begin{itemize}
    \item[-] $\mathbb{E}_{t-1}\left(\boldsymbol{X}_t\mid \mathcal{F}_{t-1}\right)\equiv \mu_t = 0$, 
    \item[-] $\mathbb{E}_{t-1}\left(\boldsymbol{X}_t \boldsymbol{X}_t^{\prime}\mid \mathcal{F}_{t-1}\right) \equiv H_t$.
\end{itemize}
For our purposes, the stochastic process $\boldsymbol{X}_t$
represents asset returns, we assume that the asset returns dynamic is: 
\begin{equation}\label{eq:1}
   \mathbf{X}_t=H_t^{\frac{1}{2}} \eta_t \text { with } \eta_t \mid \mathcal F_{t-1}\sim \mathrm{N}\left(0, I_p\right) . 
\end{equation}
Consider two generic random variables, $X_{i,t}$ and $X_{j,t}$, their correlation is: 
\begin{equation}
\rho_{ij,t}=\frac{\mathbb{E}_{t-1}({X}_{i,t} {X}_{j,t})}{\sqrt{\mathbb{E}_{t-1}({X}^2_{i,t})\mathbb{E}_{t-1}({X}^2_{j,t}})}.
\end{equation}
We decompose the covariance matrix $H_t$ to extrapolate $R_t$, the correlation of $\boldsymbol{X}_t$: 
\begin{equation}\label{decomposition}
    H_t = D_t R_t D_t, \quad \text{ where: } D_t^2 = \text{diag} (H_t).
\end{equation}
To model $R_t$,  \citet{engle2002dynamic} assumes that $\boldsymbol{X}_t \mid \mathcal{F}_{t-1}\sim \mathrm{N}(0,H_t)$, 
where the residuals $\varepsilon_t = D_t ^{-1} \boldsymbol{X}_t$ are such that $\mathbb{E}\left(\varepsilon_t \varepsilon_t^{\prime}\mid \mathcal{F}_{t-1}\right)=R_t$ is the  correlation matrix of the returns at time $t$, modeled according to the following linear transformation:
\begin{equation}\label{nonlinear_transformation}
R_t=\operatorname{diag}\left(q_{11, t}^{-\frac{1}{2}} \ldots q_{p p, t}^{-\frac{1}{2}}\right) Q_t \operatorname{diag}\left(q_{11, t}^{-\frac{1}{2}} \ldots q_{p p, t}^{-\frac{1}{2}}\right),
\end{equation}
where  \( \Bar{q}_{ij} \simeq \Bar{\rho}_{ij} \), and \( \rho_{ij} \) is the element \( i,j \) of the correlation matrix of returns:
\begin{equation}
    \rho_{ij,t}=\frac{q_{ij,t}}{\sqrt{q_{ii,t}q_{jj,t}}}.
\end{equation}
The Dynamic Conditional Correlation (DCC) model is an autoregressive process given by:
\begin{equation}\label{dcc}
    Q_t = \Bar{Q}(1-\alpha-\beta)+\alpha(\varepsilon_{t-1}\varepsilon_{t-1}^{\prime})+\beta Q_{t-1},
\end{equation}
where \( \Bar{Q} \) is the target matrix, and \( \alpha \) and \( \beta \) are non-negative parameters satisfying \( \alpha + \beta < 1 \).
From the residuals \( \varepsilon_{i,t} = D^{-1}_{i,t} x_{i,t} = \frac{x_{i,t}}{\sqrt{h_{ii,t}}} \),  the estimation of the volatility of the asset return $h_{ii,t}$ over time for each asset \( i \) is obtained by fitting GARCH-type models.

In large dimensions, the estimation of the unconditional correlation matrix $\Bar{Q}$ suffers from overfitting due to the curse of the dimensionality problem.
\citet{engle2019large} solve this issue with the DCC-NL model, introducing a non-linear shrinkage of the eigenvalues, $\Bar{Q}$, using the QuEST function in \citet{ledoit2015spectrum} that produces a more stable estimate of the matrix. 

We propose an extension of the DCC model by relaxing the assumption of normally distributed returns in favor of the more flexible nonparanormal distribution. To capture the evolving dependence structure among asset returns, we employ nonparametric measures of association, specifically Spearman's rho and Kendall's tau.
\section{The Dynamic Conditional SKEPTIC Model} \label{section_3}
\subsection{Nonparanormal}

Let ${\boldsymbol{X}_t}$ denote the demeaned $p$-dimensional vector of asset returns,
so that the conditional covariance matrix is $H_t = D_t R_t D_t$,
where $D_t$ collects the univariate conditional volatilities estimated via GARCH-type processes.
We define the standardized residuals
$
{\varepsilon}_t = D_t^{-1}\boldsymbol{X}_t
= (\varepsilon_{1t}, \ldots, \varepsilon_{pt})^{\prime}
$.
To relax the assumption of multivariate normality, we assume that the standardized residuals follow a 
{nonparanormal} distribution, so we consider 
 ${f}=\{f_1,\ldots,f_p\}$,  a set of strictly increasing univariate functions, such that for each margin $j=1,\ldots,p$, we define
$
\nu_{t,j} = f_j(\varepsilon_{t,j}).
$
The component-wise transformation $\nu_t=f\left(\varepsilon_t\right)=\left(\nu_{1,t}, \ldots, \nu_{p,t}\right)^{\prime}$ follows a multivariate normal distribution with zero mean and correlation matrix $R_t$:
$
\nu_t \mid \mathcal{F}_{t-1} \sim \mathrm{N}\left(0, R_t\right).
$
Under the nonparanormal specification, the observed returns satisfy $X_t=D_t\varepsilon_t$, while the latent Gaussian scores are given by $\nu_t=f(\varepsilon_t)$. Defining $Z_t=D_t\nu_t$, we have
\(
Z_t\mid\mathcal F_{t-1}\sim \mathrm{N}(0,H_t).
\)
Since \(Z_t\) represents the Gaussian transformation of \(X_t\) on the original conditional scale, the conditional nonparanormal density of \(X_t\) is given by:
\begin{equation}\label{eq:npn_pdf}
f_{\boldsymbol{X_t}}(x_t)
= \frac{1}{(2 \pi)^{p/2} |H_t|^{1/2}}
  \exp\!\left\{
    -\frac{1}{2} f(x_t)' H_t^{-1} f(x_t)
  \right\}
  \prod_{j=1}^p |f_j'(x_{t,j})|,
\end{equation}
we stress the fact that $f(x_t) = z_t = D_t f(\epsilon_t)$, and $\varepsilon_t = D^{-1}_t x_t$. 
In this setting, the conditional covariance structure $H_t$ captures the dynamic evolution
of the second moments, while the nonlinear transformations $f_j$ account for departures from normality
in the marginal distributions.
Importantly, the dependence structure encoded in $R_t$ is not defined directly on the raw variables
$\boldsymbol{X}_t$, but rather on their component-wise transformations.

Let $F_j$ denote the continuous marginal distribution function of the
$j$th standardized residual $\varepsilon_{t,j}$. By Sklar's theorem, the
joint distribution of $\boldsymbol{\varepsilon}_t$ can be represented through
a Gaussian copula. Specifically, define the probability integral transforms
\(
U_{j,t}
=
F_j(\varepsilon_{j,t}),
j=1,\ldots,p,
\)
and let
\(
\boldsymbol{U}_t
=
(U_{1,t},\ldots,U_{p,t})^{\prime}.
\)
Conditional on $\mathcal{F}_{t-1}$, the dependence structure of
$\boldsymbol{\varepsilon}_t$ is described by the Gaussian copula
\[
C_{\boldsymbol{R}_t}(u_1,\ldots,u_p)
=
\Phi_{\boldsymbol{R}_t}
\left(
\Phi^{-1}(u_1),
\ldots,
\Phi^{-1}(u_p)
\right),
\qquad
(u_1,\ldots,u_p)\in[0,1]^p,
\]
where $\Phi_{\boldsymbol{R}_t}$ denotes the cumulative distribution function
of a $p$-dimensional standard Gaussian random vector with correlation matrix
$\boldsymbol{R}_t$, and $\Phi^{-1}$ denotes the standard normal quantile
function.

Accordingly, the conditional joint distribution of
$\boldsymbol{\varepsilon}_t$ satisfies
\[
\Pr\!\left(
\varepsilon_{t,1}\leq x_1,\ldots,\varepsilon_{t,p}\leq x_p
\mid
\mathcal{F}_{t-1}
\right)
=
C_{\boldsymbol{R}_t}
\left(
F_1(x_1),
\ldots,
F_p(x_p)
\right).
\]
Thus, $\boldsymbol{R}_t$ is the conditional correlation matrix of the latent
Gaussian scores and parametrizes the time-varying Gaussian copula, rather than
the Pearson correlation matrix of the standardized residuals.

Since the marginal distribution functions $F_j$ are unknown, we estimate them
nonparametrically using the rescaled empirical cumulative distribution
functions
\[
\widehat{F}_j(w)
=
\frac{1}{T+1}
\sum_{s=1}^{T}
\mathbb{I}\!\left(
\varepsilon_{j,s}\leq w
\right),
\qquad
j=1,\ldots,p.
\]
The corresponding pseudo-observations are
\(
\widehat{U}_{j,t}
=
\widehat{F}_j(\varepsilon_{j,t}),
\)
from which we construct the nonparametric latent Gaussian scores
\(
\widehat{\nu}_{j,t}
=
\Phi^{-1}\!\left(
\widehat{U}_{j,t}
\right)
=
\Phi^{-1}\!\left(
\widehat{F}_j(\varepsilon_{j,t})
\right).
\)
Collecting these scores gives
\(
\widehat{\boldsymbol{\nu}}_t
=
\left(
\widehat{\nu}_{1,t},
\ldots,
\widehat{\nu}_{p,t}
\right)^{\prime}.
\)
At the population level, the corresponding latent vector satisfies
\[
\boldsymbol{\nu}_t
=
\left(
\Phi^{-1}(F_1(\varepsilon_{1,t})),
\ldots,
\Phi^{-1}(F_p(\varepsilon_{p,t}))
\right)^{\prime},
\qquad
\boldsymbol{\nu}_t
\mid
\mathcal{F}_{t-1}
\sim
{N}
\left(
\boldsymbol{0},
\boldsymbol{R}_t
\right).
\]

\cite{fan2023estimation} and \cite{cordoni2024consistent} extend the nonparanormal framework to dependent data,
proposing vector autoregressive structures for latent Gaussian scores.
In our setting, we focus instead on modeling the second-order dynamics through $R_t$,
leading to the Dynamic Conditional SKEPTIC specification.
\subsection{Dynamic Conditional SKEPTIC}

{Under the nonparanormal assumption, we estimate the long-run correlation
target $\Bar{\mathcal{Q}}$ entering the DCC recursion in \eqref{dcc} using
the rank-based dependence measures Spearman's rho and Kendall's tau.
The resulting SKEPTIC estimator is then plugged into the autoregressive
correlation recursion to obtain the time-varying conditional correlation
matrix.}

Let
\(
\boldsymbol{\nu}_t
=
\boldsymbol{f}(\boldsymbol{\varepsilon}_t)
=
\left(
f_1(\varepsilon_{1t}),
\ldots,
f_p(\varepsilon_{pt})
\right)^{\prime}
\)
denote the latent Gaussian vector, and define
\(
\Bar{\mathcal{Q}}
\simeq
\mathbb{E}
\left(
\boldsymbol{\nu}_t
\boldsymbol{\nu}_t^{\prime}
\right),
\) the target matrix that starts the Dynamic Conditional SKEPTIC recursion. 
Since the components of $\boldsymbol{\nu}_t$ have unit variance,
$\Bar{\mathcal{Q}}$ is the
long-run correlation matrix of the latent Gaussian process.

\begin{Lemma}
\label{kruskal}
Assume that the latent Gaussian vector has long-run correlation matrix
$\Bar{\mathcal{Q}}=(\Bar q_{ij})$. Then, for every $i\neq j$,
\[
\Bar q_{ij}
=
2\sin\left(
\frac{\pi}{6}\rho_{ij}^{s}
\right)
=
\sin\left(
\frac{\pi}{2}\tau_{ij}
\right),
\]
where $\rho_{ij}^{s}$ and $\tau_{ij}$ denote, respectively, the
population Spearman's rho and Kendall's tau associated with the
$i$th and $j$th components of the latent process.
\end{Lemma}

Given the observations
$\{\boldsymbol{\nu}_t\}_{t=1}^{T}$, the sample Kendall's tau is defined
as
\[
\widehat{\tau}_{ij}
=
\frac{2}{T(T-1)}
\sum_{1\leq s<s'\leq T}
\operatorname{sign}
\left(
\nu_{i,s}-\nu_{i,s'}
\right)
\operatorname{sign}
\left(
\nu_{j,s}-\nu_{j,s'}
\right),
\]
while the sample Spearman's rho is
\[
\widehat{\rho}_{ij}^{s}
=
\frac{
\sum_{s=1}^{T}
(\zeta_{i,s}-\Bar{\zeta}_i)
(\zeta_{j,s}-\Bar{\zeta}_j)
}{
\sqrt{
\sum_{s=1}^{T}
(\zeta_{i,s}-\Bar{\zeta}_i)^2
}
\sqrt{
\sum_{s=1}^{T}
(\zeta_{j,s}-\Bar{\zeta}_j)^2
}
},
\]
where $\zeta_{i,s}$ and $\zeta_{j,s}$ are the ranks of
$\nu_{i,s}$ and $\nu_{j,s}$ in their respective samples, and
\[
\Bar{\zeta}_i
=
\Bar{\zeta}_j
=
\frac{T+1}{2}.
\]

\begin{Proposition}[SKEPTIC plug-in estimator of
$\Bar{\mathcal{Q}}$]
\label{prop:qbar_skeptic}
The SKEPTIC plug-in estimators of the long-run latent correlation
matrix $\Bar{\mathcal{Q}}$ is defined as follows. Let us consider the $(i,j)$ element, the resulting matrices are: 
\[
[\widehat{\Bar{\mathcal{Q}}}^{\,\tau}]_{ij} = \widehat{\Bar q}_{ij}^{\,\tau}
=
\begin{cases}
\displaystyle
\sin\left(
\frac{\pi}{2}\widehat{\tau}_{ij}
\right),
& i\neq j,\\[1ex]
1,
& i=j,
\end{cases}
\]
and
\[
[\widehat{\Bar{\mathcal{Q}}}^{\,\rho}]_{ij} = \widehat{\Bar q}_{ij}^{\,\rho}
=
\begin{cases}
\displaystyle
2\sin\left(
\frac{\pi}{6}\widehat{\rho}_{ij}^{s}
\right),
& i\neq j,\\[1ex]
1,
& i=j.
\end{cases}
\]
Thus, either
$\widehat{\Bar{\mathcal{Q}}}^{\,\tau}$ or
$\widehat{\Bar{\mathcal{Q}}}^{\,\rho}$ can be used as a rank-based
plug-in estimator of the long-run target matrix in the DCC recursion.
\end{Proposition}

\begin{Definition}
\label{dcs}
The Dynamic Conditional \texttt{SKEPTIC} model is defined in
\eqref{dcc}, where the time-varying conditional correlation
matrix is
\begin{equation}
\mathcal{R}_t
=
\operatorname{diag}\left(
q_{11,t}^{-\frac{1}{2}},
\ldots,
q_{pp,t}^{-\frac{1}{2}}
\right)
\mathcal{Q}_t
\operatorname{diag}\left(
q_{11,t}^{-\frac{1}{2}},
\ldots,
q_{pp,t}^{-\frac{1}{2}}
\right),
\end{equation}
and the $p\times p$ symmetric positive-definite matrix
$\mathcal{Q}_t=(q_{ij,t})$ follows the recursion
\begin{equation}
\label{ff}
\mathcal{Q}_t
=
(1-\alpha-\beta)
\widehat{\Bar{\mathcal{Q}}}
+
\alpha
\boldsymbol{\nu}_{t-1}
\boldsymbol{\nu}_{t-1}^{\prime}
+
\beta
\mathcal{Q}_{t-1},
\end{equation}
where $\widehat{\Bar{\mathcal{Q}}}$ denotes either
$\widehat{\Bar{\mathcal{Q}}}^{\,\tau}$ or
$\widehat{\Bar{\mathcal{Q}}}^{\,\rho}$, as defined in
Proposition~\ref{prop:qbar_skeptic}. The parameters $\alpha$ and
$\beta$ are non-negative scalars satisfying
\[
\alpha+\beta<1.
\]
\end{Definition}
For every $t$, the latent Gaussian vector satisfies
\(
\boldsymbol{\nu}_t
\mid
\mathcal{F}_{t-1}
\sim \mathrm{N}
\left(
\boldsymbol{0},
\mathcal{R}_t
\right),
\)
so that $\mathcal{R}_t$ remains the conditional correlation matrix of
the latent Gaussian scores.
The DCS extends the DCC specification by replacing the conventional
estimator of the long-run correlation target with the rank-based
SKEPTIC estimator. The matrix
$\widehat{\Bar{\mathcal{Q}}}$ provides the static SKEPTIC target,
whereas the arrival of new information through
$\boldsymbol{\nu}_{t-1}\boldsymbol{\nu}_{t-1}^{\prime}$ and the
autoregressive term $\mathcal{Q}_{t-1}$ generates the time variation in
$\mathcal{Q}_t$ and $\mathcal{R}_t$. Since the conditional latent
Gaussian assumption is preserved at every date, each
$\mathcal{R}_t$ inherits the dependence structure identified by the
SKEPTIC target, while evolving dynamically through
\eqref{ff}. Thus, time variation is generated by the DCC recursion and
does not require Kendall's tau or Spearman's rho to be defined as
time-varying sample statistics.
\begin{remark}
\label{rem:skeptic-transformations}
Although the SKEPTIC estimator is introduced through the latent Gaussian
process, its implementation does not require explicit estimation of the
unknown marginal transformation functions, as highlighted in \cite{liu2012high}. Spearman's rho and Kendall's
tau are rank-based statistics and are therefore invariant under strictly
increasing marginal transformations. Consequently, the rank-based
correlation target can be computed directly from the standardized
residuals. 
\end{remark}

\begin{remark}[\cite{aielli2013dynamic}]
In the standard DCC specification, the equality
\(
\Bar{\mathcal{Q}}
=
\mathbb{E}\!\left(
\boldsymbol{\nu}_t\boldsymbol{\nu}_t^{\prime}
\right)
\)
does not generally hold because of the normalization used to obtain
$\mathcal{R}_t$. Following the DCC correction of \citet{aielli2013dynamic}, we can consider the corrected DCS
specification as
\[
\mathcal{Q}_t
=
(1-\alpha-\beta)\Bar{\mathcal{Q}}
+
\alpha
\mathcal{Q}_{t-1}^{*1/2}
\boldsymbol{\nu}_{t-1}
\boldsymbol{\nu}_{t-1}^{\prime}
\mathcal{Q}_{t-1}^{*1/2}
+
\beta\mathcal{Q}_{t-1},
\]
where
\(\mathcal{Q}_t^{*}
=
\operatorname{diag}(\mathcal{Q}_t).
\)
The corresponding long-run target satisfies
\(
\Bar{\mathcal{Q}}
=
\mathbb{E}\!\left[
\mathcal{Q}_t^{*1/2}
\boldsymbol{\nu}_t
\boldsymbol{\nu}_t^{\prime}
\mathcal{Q}_t^{*1/2}
\right].
\)
The corrected version can be implemented by applying the
SKEPTIC estimator to the transformed process
\(
\boldsymbol{\nu}_t^{*}
=
\mathcal{Q}_t^{*1/2}\boldsymbol{\nu}_t.
\)
\end{remark}
The baseline DCS follows the conventional DCC targeting specification, whereas the corrected recursion provides the theoretically consistent targeting formulation of \cite{aielli2013dynamic}.
\section{Theoretical Properties}\label{section_4}
We establish the theoretical foundation of the SKEPTIC estimator for an autoregressive process, demonstrating that this novel approach is a reliable extension to the DCC model. 
We start recalling the stationarity conditions outlined in the work of \citet{fermanian2017stationarity}, where the DCC is reformulated in terms of a Markov chain. In our case, under the assumption of  nonparanormal distribution, we define the following model:
\[
\begin{aligned}
&W_t := \left(W_t^{(1)}, W_t^{(2)}, W_t^{(3)}, W_t^{(4)}\right)^{\prime} \quad \text{where} \\[0.5em]
&W_t^{(1)} := \left(\operatorname{Vecd}\left(D_t\right)\right)^{\prime}, 
& &W_t^{(2)} := \left(\vec{X}_t\right)^{\prime}, \ \text{where } \vec{X}_t = (X_{1 t}^2, \hdots, X_{p t}^2)^{\prime}, \\[0.5em]
&W_t^{(3)} := \left(\operatorname{Vech}\left(\mathcal{Q}_t\right)\right)^{\prime}, 
& &W_t^{(4)} := \left(\operatorname{Vech}\left(\nu_t \nu^{\prime}_t\right)\right)^{\prime}.
\end{aligned}
\]
The dimensions of the four previous random vectors are, respectively, $1 \times p, 1 \times p, 1 \times p(p+1)/2$, and $1 \times p(p+1)/2$. Their sum,  dimension of $W_t$, is denoted by $d$. Simple block matrix calculations show that there exist random matrices, $\left(T_t\right)$, and a vector process, $\left(\kappa_t\right)$, such that the dynamics of $W_t$  %- \textcolor{red}{any solution of the DCC model} - 
may be rewritten as:
\begin{equation}\label{Dcc_mc}
  W_t=T_t \cdot W_{t-1}+\kappa_t,  
\end{equation}
for any $t$, where $T:=[T_{ij,t}]_{1\leq i,j\leq 4}$ is a block matrix.\footnote{We refer to  \citet{fermanian2017stationarity}, where $\mu=\nu=r=s=1$.} 
To determine the stationarity of the DCS process, we  define the data-generating process of $W_t$:
\begin{equation}\label{delta_t}
    \delta_t := \mathcal{R}_t^{-\frac{1}{2}}\nu_t = \mathcal{R}_t^{-\frac{1}{2}} D_t^{-\frac{1}{2}} W_t.
\end{equation}
Note that $\mathbb{E}\left[\delta_t\mid \mathcal{F}_{t-1}\right]=0$ and $\mathbb{E}\left[\delta_t \delta_t^{\prime}\mid \mathcal{F}_{t-1}\right]=I_p$ by construction. The definition of these innovations implies that, for every $t, \sigma\left(\delta_j, j \leq t\right) \subset \sigma\left(\nu_j, j \leq t\right) \subset \mathcal{F}_t$, where $\sigma$ is the sigma algebra.
\begin{Proposition}(Stationarity of DCC (\citealp{fermanian2017stationarity}))\label{stationarity}
    Given the process $\delta_t$ in \eqref{delta_t}, we state that:
    \begin{itemize}
        \item[(A0)] $\left(\delta_t\right)_{t \in \mathbb{Z}}$ possesses the Markov property with respect to the filtration $\mathcal{F}$. In particular, $\mathbb{E}\left[\delta_t \mid \mathcal{F}_{t-1}\right]=E\left[\delta_t \mid W_{t-1}\right]$ for every $t$.
        \item[(A1)]\[
\left\|M_1\right\|_{\infty}+\left\|N_1\right\|_{\infty}<1 
\quad \text{and} \quad 
\left\|B\right\|_{\infty}<1.
\]
where: $Vecd(D_t)= V_0+M_1 Vecd(D_{t-1})+N_1 \vec{W}_t$ are the univariate GARCH(1,1) models, and the matrix $B\geq 0$, in other words, all the elements must be non-negative. 
         \item[(A2)] For some 
$\mathbb{E}\left[\left\|\delta_t\right\|^{2}\right]<\infty$ and $\rho\left(T^*\right)<1$, where $T^*:=\sup _{\mathbf{w} \in \mathbb{R}^d} E\left[\left|T_t\right| \mid W_{t-1}=\mathbf{w}\right]$, for any norm $\|\cdot\|$.
        \item[(A3)] The law of $\delta_t$ given that $W_{t-1}=\mathbf{w}$ is continuous concerning the Lebesgue measure, and its density is denoted by $f_{\delta_t}(\cdot \mid \mathbf{w})$, for every $\mathbf{w} \in \mathbb{R}^d$ and $t$. The function $\mathbf{w} \mapsto f_{\delta_t}(\delta \mid \mathbf{w})$ is continuous for every $\delta \in \mathbb{R}^p$ and $t$. There exists an integrable function $H$ such that $\sup _t \sup _{\mathbf{w} \in \mathbb{R}^d} f_{\delta_t}(\delta \mid \mathbf{w}) \leq H(\delta)$ for every $\delta \in \mathbb{R}^p$. Moreover, $\sup _t E\left[\left\|\delta_t\right\|^{2 } \mid W_{t-1}=\mathbf{w}\right] \leq \Bar{h}(\|\mathbf{w}\|)$, for some function $\Bar{h}$ that satisfies $\lim _{v \rightarrow+\infty} \Bar{h}(v) / v^\gamma=0$ for every $\gamma>0$.
    \end{itemize}
   Under the assumptions (A0, A1, A2, A3), the process $(W_t, D_t, \mathcal{R}_t)$ is  stationary. 
\end{Proposition}

When the assumptions are satisfied, the correlation matrix estimated with the Dynamic Conditional SKEPTIC in Definition \ref{dcs} is an autoregressive process, which is a stationary Markov chain.

\subsection{Mixing Conditions of Semiparametric Gaussian Process}
Under the assumption of $\boldsymbol{X}_t\mid \mathcal{F}_{t-1} \sim \operatorname{NPN}(0,H_t,f)$. For each sequence of random variables, we derive its mixing properties. %exploiting the results of \citet{beare2010copulas}. 
\begin{Definition}(\citealp{beare2010copulas})\label{def_beta_mixing}
 The $\beta$-mixing coefficients $\left\{\beta_k: k \in \mathbb{N}\right\}$ that correspond to the sequence of random variables $\left\{X_t\right\}$ are given by
$$
\beta_k=\frac{1}{2} \sup _{m \in \mathbb{Z}} \sup _{\left\{A_i\right\},\left(B_j\right\}} \sum_{i=1}^I \sum_{j=1}^J\left|P\left(A_i \cap B_j\right)-P\left(A_i\right) P\left(B_j\right)\right|
$$
where the second supremum is taken over all finite partitions $\left\{A_1, \ldots, A_I\right\}$ and $\left\{B_1, \ldots, B_J\right\}$ of $\Omega$ such that $A_i \in \mathcal{F}_{-\infty}^m$ for each $i$ and $B_j \in \mathcal{F}_{m+k}^{\infty}$ for each $j$.
\end{Definition}
\begin{Proposition}\label{proposition_beare}
    The semiparametric Gaussian copula $C_{R_t}$ is symmetric and continuous with square-integrable density $c$ with $\rho_C<1$, then 
    there exists $A<\infty$ and $\gamma>0$ such that $\beta_k \leq A e^{-\gamma k}$ for all $k$. 
\end{Proposition}
The proof is in the Appendix \ref{proof_beare}. 
Moreover, by Proposition 2, and 4 in \citet{remillard2012copula}, the Gaussian copula process is also a $\rho$-mixing process, where the $\rho$-mixing coefficients are defined as follows. 
\begin{Definition}(\citealp{beare2010copulas})
    The $\rho$-mixing coefficients $\left\{\rho_k: k \in \mathbb{N}\right\}$ that correspond to the sequence of random variables $\left\{X_t\right\}$ are given by
$$
\rho_k=\sup _{m \in \mathbb{Z}} \sup _{f, g}|\operatorname{Corr}(f, g)|,
$$ 
where the second supremum is taken over all square-integrable random variables $f$ and $g$ measurable for $\mathcal{F}_{-\infty}^m$ and $\mathcal{F}_{m+k}^{\infty}$ respectively, with positive and finite variance, and where $\operatorname{Corr}(f, g)$ denotes the correlation between $f$ and $g$. 
\end{Definition}

The semiparametric Gaussian copula process exhibits strong mixing properties, implying that serial dependence among the observations decreases over time. Assuming that stationarity and mixing conditions hold for the process, concentration inequalities can be obtained for the estimated matrices using the DCS process. 

Under stationarity and mixing conditions, we know from \cite{han2018exponential} that defining $m\in\{\tau,\rho\}$, and let $\widehat{\Bar Q}^{\,m}$ denote the SKEPTIC estimator based on Kendall's tau or Spearman's rho, respectively. The plug-in matrices have the following rate of convergence 
\[
\left\|
\widehat{\Bar Q}^{\,m}-\Bar Q^{\,m}
\right\|_{\max}
=
O_p\left(
\sqrt{\frac{\log(Tp)}{T}}
\right).
\]

Since $\widehat{\Bar Q}^{\,m}$ enters the DCS recursion as the long-run target, it remains to determine how this estimation error propagates through the dynamics of $Q_t^m$ and through the normalization used to obtain the conditional correlation matrix $R_t^m$.

\begin{Proposition}\label{dcs_error}
Let $m\in\{\tau,\rho\}$, and let $\widehat{\bar Q}^{\,m}$ denote the SKEPTIC estimator of the long-run target matrix based on Kendall's tau when $m=\tau$ and on Spearman's rho when $m=\rho$, as defined in Lemma~\ref{kruskal}.

Define the population recursion
\[
Q_t^m
=
(1-\alpha-\beta)\bar Q^m
+
\alpha\nu_{t-1}\nu_{t-1}^{\prime}
+
\beta Q_{t-1}^m,
\]
and its estimated counterpart
\[
\widehat Q_t^m
=
(1-\alpha-\beta)\widehat{\bar Q}^{\,m}
+
\alpha\nu_{t-1}\nu_{t-1}^{\prime}
+
\beta\widehat Q_{t-1}^m.
\]

Suppose that Propositions~\ref{stationarity} and~\ref{proposition_beare} hold and that
\[
\left\|
\widehat{\bar Q}^{\,m}-\bar Q^m
\right\|_{\max}
=
O_p\left(
\sqrt{\frac{\log(Tp)}{T}}
\right).
\]
Assume that the two recursions are initialized at their respective long-run targets,
\[
Q_0^m=\bar Q^m,
\qquad
\widehat Q_0^m=\widehat{\bar Q}^{\,m},
\]
and that the diagonal elements of $Q_t^m$ and $\widehat Q_t^m$ are uniformly bounded away from zero.

Then
\[
\left\|
\widehat R_t^m-R_t^m
\right\|_{\max}
=
O_p\left(
\sqrt{\frac{\log(Tp)}{T}}
\right).
\]
\end{Proposition}

Proof of Proposition \ref{dcs_error} is in Appendix \ref{proof_dcs_error}.
\section{Estimation}\label{section_5}
The parameter estimation procedure in large dimensions is burdensome due to the difficulty in inverting the matrix $H_t$ in the log-likelihood of $z_t \equiv f(x_t) \mid \mathcal{F}_{t-1} \sim \operatorname{N}\left(0, D_t \mathcal{R}_t D_t\right)$:
\begin{align}
    L = \sum_{t=1}^T  \left( \frac{1}{2}\log\left|H_t\right|+\frac{1}{2}z_t^\prime H_t^{-1}z_t-\sum_{j=1}^p\log\left(\left|\frac{\partial z_{tj}}{\partial x_{tj}}\right|\right)\right),\notag
\end{align}
which is proportional to: 
\begin{align}
L \propto
\frac{1}{2}\sum_{t=1}^T\left( \log\left|H_t\right|+z_t^\prime H_t^{-1}z_t\right).
\end{align}
We rewrite the function to separate the correlation and volatility components: 
\begin{align}\label{likelihood}\notag
    L &=  \frac{1}{2} \sum_{t=1}^T\left( \log\left|H_t\right|+z_t^\prime H_t^{-1}z_t\right),\\\notag
   % &=\frac{1}{2} \sum_{t=1}^T\left(\log\left|D_t \mathcal{R}_t D_t\right|+z_t^\prime D_t^{-1} \mathcal{R}_t^{-1} D_t^{-1} z_t\right),\\\notag
    &=\frac{1}{2}\sum_{t=1}^T\left( \log\left|
    D_t\right|^2+ \log\left|\mathcal{R}_t\right|+ \nu_t^{\prime} \mathcal{R}_t^{-1} \nu_t\right) \quad \text{where: } \nu_t = f(\varepsilon_t),\\\notag
   & = \frac{1}{2}\sum_{t=1}^T\left( \log\left|D_t\right|^2+z_t^{\prime}D^{-1}_t D^{-1}_t z_t-\nu_t^{\prime} \nu_t+ \log\left|\mathcal{R}_t\right|+ \nu_t^{\prime} \mathcal{R}_t^{-1} \nu_t\right).
\end{align}
Denote with $\theta$ the parameters in $D_t$  and with $\phi$ the parameters in $\mathcal{R}_t$. The log-likelihood function can be expressed as the sum of a volatility component and a correlation component:
$$
L(\theta, \phi)=L_V(\theta)+L_C(\theta, \phi) .
$$
where the volatility term is
\begin{equation}\label{volatilities}
    L_V(\theta)=\frac{1}{2} \sum_{t=1}^T\left(\log \left|D_t\right|^2+z_t^{\prime} D_t^{-2} z_t\right),
\end{equation}

and the correlation component is
\begin{equation}
  L_C(\theta, \phi)=\frac{1}{2} \sum_{t=1}^T\left(\log \left|\mathcal{R}_t\right|+\nu_t^{\prime} \mathcal{R}_t^{-1} \nu_t-\nu_t^{\prime} \nu_t\right). 
\end{equation}
Fitting individual GARCH-type models minimizes the likelihood function for the volatility component.
%\begin{equation}
 % L_V(\theta)=\frac{1}{2} \sum_{t=1}^T \sum_{i=1}^n\left(\log \left(h_{i, t}\right)+\frac{z_{i, t}^2}{h_{i, t}}\right),  
%\end{equation}
Once the volatility parameters are determined, we need to perform the second step of the estimation procedure, which requires minimizing $\phi$ given $\widehat{\theta}$:
\begin{equation}
  \min _\phi\left\{L_C(\widehat{\theta}, \phi)\right\}  .
\end{equation}
The estimation of $\phi=(\alpha, \beta)$  proposed in \citet{engle2002dynamic} leads to major issues as dimensionality grows. This difficulty aligns with the curse of dimensionality described in \citet{engle2019large}. Specifically, as the concentration ratio $p/T$ increases, the estimation of $\mathcal{R}_t$ suffers from overfitting, and when $p/T$ is equal to 1, the matrix $\mathcal{R}_t$ becomes singular and the procedure inapplicable.

We follow \citet{pakel2021fitting} and use the composite likelihood function that estimates the parameters by summing the quasi-likelihoods of asset subsets, thereby addressing lower-dimensional marginal densities.
This method involves determining the conditional variances with a GARCH-type model and then fitting a likelihood based on correlation $2 \times 2$. More precisely, once the values of $\widehat{\theta}$ that maximize \eqref{volatilities} are obtained (\citealp{fermanian2024model}), we estimate the parameter set $\phi$ maximizing:
\begin{equation}
    L_{C,2T}(\widehat{\theta}, \phi)=-\frac{1}{T} \sum_{t=1}^T \sum_{k=1}^\mathcal{L}\left[\log \left(\left|\mathcal{R}_t^{(k)}\right|\right)+\nu_t^{(k) \prime}\left(\mathcal{R}_t^{(k)}\right)^{-1} \nu_t^{(k)}\right],
\end{equation}
where $R_t^{(k)}$ is generated by the DCS recursion, whose long-run target is estimated using SKEPTIC, with $k$ corresponding to a prespecified pair of indices in $\{1, \ldots, N-1\}$,  $\nu_t^{(k)}$ is the corresponding $2 \times 1$ sub-vector of the standardized residuals $\nu_t=D_t^{-1} z_t$. We choose contiguous pairs $\nu_t^{(1)}=\left(\nu_{1 t}, \nu_{2 t}\right)^{\prime}, \nu_t^{(2)}=$ $\left(\nu_{2 t}, \nu_{3 t}\right)^{\prime}, \ldots, \nu_t^{(N-1)}=\left(\nu_{(N-1) t}, \nu_{N t}\right)^{\prime}$. %so that $L=\mathrm{N}(N-1)/2$. 
By restricting attention to $2\times 2$ matrices, we avoid the need to invert large-dimensional correlation matrices, reducing both computational complexity and the risk of estimation inaccuracies.

\subsection{Asymptotic Properties of Composite Likelihood Estimator}
Let us recall a bivariate DCS model for $x_t \mid \mathcal{F}_{t-1} \sim \operatorname{NPN}(0,H_t)$: 
\begin{align}
    &\mathcal{Q}_t = (1-\alpha-\beta) \Bar{\mathcal{Q}} + \alpha (\nu_{t-1} \nu_{t-1}^{\prime}) + \beta \mathcal{Q}_{t-1},
\end{align}
where: 
$$
X_{j t}=\left\{x_{1 j t}, x_{2 j t}\right\}, \quad \operatorname{Cov}\left(X_{j t} \mid \mathcal{F}_{t-1}\right)=\left(\begin{array}{cc}
h_{1 j t}^{1 / 2} & 0 \\
0 & h_{2 j t}^{1 / 2}
\end{array}\right) \mathcal{R}_{j t}\left(\begin{array}{cc}
h_{1 j t}^{1 / 2} & 0 \\
0 & h_{2 j t}^{1 / 2}
\end{array}\right),
$$
where $\mathrm{Var}\left(x_{ljt}\mid \mathcal{F}_{t-1},\eta_{lt}\right)=h_{ljt}$,  $l = 1,2$,
 $\mathrm{Cor}\left(\nu_{jt}\mid \mathcal{F}_{t-1}\right)=\mathcal{R}_{jt}$,  $\mathcal{R}=\left\{R_{jt}^{\rho}\vee R_{jt}^{\tau}\right\}$ and  $\theta_{j}=(\zeta^{\prime}_{1j},\zeta^{\prime}_{2j})^\prime$ are estimated with a GARCH-type model.
We estimate the parameters: $\phi = (\alpha, \beta)$  and the matrix $\Bar{\mathcal{Q}}$  with SKEPTIC in Lemma \ref{kruskal}.
The generic framework for performing the composite likelihood estimation procedure is the following:
\begin{align}
  \widehat{\theta}=\arg \max _{\phi \in \Phi} \sum_{j=1}^N \sum_{t=1}^T \ell_{j t}\left(\theta, \widehat{\theta}_j(\phi)\right) \quad \text { and } \quad \widehat{\lambda}_j(\theta)=\arg \max _{\theta_j \in \Theta_j} \sum_{t=1}^T \ell\left(\phi, \theta_j ; X_{j t}\right).
\end{align}
\citet{pakel2021fitting} demonstrate that the composite likelihood estimator yields consistent and asymptotically normal estimates. %under  assumptions (A1)–(A4), which are satisfied by our model.% resulting in a consistent and asymptotically normal estimator.

\section{Simulation Studies}\label{section_6}
We assess the ability of the Dynamic Conditional SKEPTIC to recover the true dependence dynamics in finite samples by comparing its parameter estimates with those obtained from DCC and DCC-NL models in a  Monte Carlo setting.
 
 We compute an empirical target matrix \(\Bar Q\) from log-returns of a set of stocks, which is used as the starting point in the correlation recursion. For each Monte Carlo replication, we simulate a \(p\)-dimensional return series \(\{x_t\}_{t=1}^T\) according to the DCC model in \eqref{dcc}, with fixed parameters $\alpha = 0.02, \beta=0.97$. Recalling that \(H_t=D_t R_t D_t\), we generate returns as \(x_t=H_t^{1/2}u_t\),  $\forall t \in \{1,\dots,T\}$.
For each asset \(i=1,\ldots,p\), the innovation \(u_{i,t}\) is drawn from a Student-\(t\) distribution \(t_\nu\!\left(0,1\right)\), and the number of degrees of freedom \(\nu\) is randomly drawn between 4 and 8. 

The empirical returns exhibit non-negligible skewness. To reproduce this feature, we introduce contamination into the simulated data. Let \(\delta\) denote the contamination level; we consider three cases, \(\delta \in \{0.01,0.05,0.10\}\), corresponding to \(1\%\), \(5\%\), and \(10\%\) of contaminated observations, respectively. The contamination is introduced by replacing, independently across assets and time, the specified proportion of the innovations \(u_{i,t}\) with draws from a \(t_\nu\!\left(0,1\right)\) distribution with \(\nu=3\), thereby increasing the number of outliers and inducing asymmetry in the simulated return distribution.

We consider sample sizes \(T \in \{500,1000,2000\}\) and cross-sectional dimensions \(p\in \{50,97,250,429\}\), in the cases: $p= 50$ and $p = 250$, we randomly select stocks from the SP500 index, the other cases ($p = 97$, $p = 429$) represent the full sample of S\&P100 and S\&P500. Once the returns are simulated, we fit the following models: DCC, DCC-NL, DCS (tau), and DCS (rho). From each fitted model we extract the parameters \((\widehat{\alpha},\widehat{\beta})\) and compare them with those used to simulate the returns.  We replicate this scheme 100 times, and the full simulation results are reported in Appendix \ref{app_simulation}, where the values are displayed by considering the different contamination levels $\delta$. 

As shown from Tables \ref{tab_sim_1}, \ref{tab_sim_2}, and \ref{tab_sim_3} in the Appendix \ref{app_simulation}, the estimates of $\alpha$ and $\beta$ are overall more stable for the DCS models with respect to the DCC-based models, ensuring robustness under non-normality. 
\section{Empirical Analysis}\label{section_7}
We conduct an empirical analysis comparing the Dynamic Conditional SKEPTIC (DCS) with the DCC model of \citet{engle2002dynamic} and the DCC-NL of \citet{engle2019large}\footnote{We thank Michael Wolf for kindly providing the codes for the DCC and DCC-NL on the website \url{https://www.econ.uzh.ch/en/people/faculty/wolf/publications.html} under the link “Programming Code.” We revisit these codes, aligning them to our purposes.}.
We download stock price data from \textit{Yahoo! Finance} of the stocks that belong to the S\&P100 and S\&P500 indices, covering the period from 02/01/2013 to 23/01/2025. After cleaning the data, the final dataset includes 429 stocks for the S\&P500 and 97 stocks for the S\&P100, each with 3034 price observations.
\subsection{Model Fit and Diagnostic Checks}

Using the full set of observations, we fit the DCC, DCC-NL, and DCS models to the S\&P100 and S\&P500 datasets. Model selection is evaluated according to the Akaike Information Criterion (AIC) and the Bayesian Information Criterion (BIC). For all the models, after having estimated the univariate volatility with a GARCH(1,1), we estimate the parameters $\phi = (\alpha, \beta)$. We remark that the estimate of the matrix ${Q}_t$ may not be positive semidefinite. In that case, we project it onto the nearest positive definite correlation matrix using the method of \citet{higham2002computing}.

In the residual analysis, we evaluate the adequacy of model fitting to the data. A well-specified econometric model should effectively filter the observed data into a white noise series, uncorrelated over time. To assess the absence of correlation, we rely on the Portmanteau test (\citealp{lutkepohl2005new}), which examines the null hypothesis that residuals exhibit no autocorrelation up to lag \(h\). The test statistic is defined as follows:
$$
\mathcal{P}_h := T \sum _{i = 1}^h (T-i)^{-1} \text{tr}(\widehat{C}_i \widehat{C}_0^{-1} \widehat{C}_i \widehat{C}_0^{-1}),
$$
where for the DCC and DCC-NL models we have:
  \[
  \widehat{C}_i = \frac{1}{T} \sum_{t = i + 1}^T \widehat{\epsilon}_t \widehat{\epsilon}_{t-i}',
  \]
and for the DCS:
  \[
  \widehat{C}_i = \frac{1}{T} \sum_{t = i + 1}^T \widehat{\nu}_t \widehat{\nu}_{t-i}'.
  \]
In large samples, and for sufficiently large \(h\), the test statistic follows an asymptotic Chi-squared distribution:
\[
\mathcal{P}_h \simeq \chi^2 _{(p^2(h-k))},
\]
where $p$ is the dimension of the residuals vector, $h$ is the number of lags considered, and $k$ is the number of estimated parameters.
Furthermore, we assess the normality of the residuals using the Jacque-Bera test based on the null hypothesis of normality, where the test statistic is derived from the third and fourth moments of the normal distribution, which asymptotically follows a \(\chi^2_2\) distribution. %The null hypothesis of the test assumes that the data follows a normal distribution.

Table \ref{Diagnostic_checks} summarizes the diagnostic checks. For the Portmanteau test, we set the number of lags $h=10$.
\begin{table}[h!]
\centering
\begin{tabular}{l|cccc|cccc}
\hline
\hline
  {} & \multicolumn{4}{c}{ \textbf{S\&P100}} & \multicolumn{4}{|c}{ \textbf{S\&P500}} \\ 
 {Model} &  \textbf{AIC}\textbf{} &  \textbf{BIC} &  \textbf{JB} &  \textbf{Port.} &  \textbf{AIC} &  \textbf{BIC} &  \textbf{JB} &  \textbf{Port.} \\ \hline
 {DCC} & -29.23 & -17.20 & $69 \times 10^{6 \text{ } ***}$ & 49.03 & -28.97 & -16.94 & $358 \times 10^{6 \text{ } ***}$ & 1071.70 \\ 
 {DCC-NL} & -29.23 & -17.20 & $67 \times 10^{6 \text{ } ***}$ & 49.14 & -28.97 & -16.93 & $361 \times 10^{6 \text{ } ***}$ & 1084.73 \\ 
 {DCS (tau)} &\textbf{-29.32} & \textbf{-17.29} & $146.73$ & 32.85 & \textbf{-29.06} & \textbf{-17.02} & $930.04$ & 655.57 \\ 
 {DCS (rho)} &-29.22&-17.18& $106.92$ & 33.43 & -29.00 & -16.93 & $549.08$ & 636.29 \\ 
 \hline
\hline
\end{tabular}
\caption{Diagnostic checks of the DCC, DCC-NL, and DCS models fitted to S\&P100 and S\&P500 dataset. JB stands for Jacque-Bera test statistic values, and Port. stands for Portmanteau test statistic values. \textit{Note:} $^* p<0.1$; $^{**} p<0.05$; $^{***} p<0.01$.}\label{Diagnostic_checks}
\end{table}
We find that the AIC and BIC suggest that our model shows a slight improvement to fit the data, the resulting residuals are uncorrelated for all the models, and in the case of DCS models they satisfy the normal distribution, as highlighted in the results of the Jacque Bera test, where the null hypothesis of normality is not rejected. 
\subsection{Portfolio Construction}
We evaluate the out-of-sample performance of our correlation matrix estimators, which are used to construct the weights of Markowitz portfolios.
%Following \cite{engle2019large}, we adopt the convention that 21 consecutive trading days constitute a "month".
Each dataset is divided into an in-sample period and an out-of-sample period. The in-sample period includes the first 1.500 observations, from 02/01/2013 to 14/12/2018. The out-of-sample period considers returns from 15/12/2018 to 23/01/2025, consisting of 73 months, where we adopt the convention that 21 consecutive trading days constitute one month.
During each month portfolio weights are fixed and no transactions occur.
We denote the investment dates by $ h = 1, \ldots, 73 $. On each investment date \( h \), the covariance matrix is estimated using the most recent $T = 1.500 $ daily returns, roughly corresponding to six years of historical data.

We estimate the global minimum variance (GMV) portfolio without short-selling constraints.  Let ${H}_t$ denote the covariance matrix at time $t$. The optimization problem is formulated as follows:
\[
\begin{aligned}
& \min_{w} \, w^{\prime} H_t w \quad 
\text{such that} & \quad w^{\prime} \mathbf{\iota} = 1,
\end{aligned}
\]
where \( \mathbf{\iota} \) is a vector of ones with dimensions \( p \times 1 \). The analytical solution is given by:
\begin{equation}\label{top}
  w_t = \frac{H_t^{-1} \mathbf{\iota}}{\mathbf{\iota}^{\prime} H_t^{-1} \mathbf{\iota}}. 
\end{equation}
In practice, the unknown \( H_t \) is replaced with an estimator \( \widehat{H}_t \), resulting in the feasible portfolio:
\begin{equation}
  \widehat{w}_t = \frac{\widehat{H}_t^{-1} \mathbf{\iota}}{\mathbf{\iota}^{\prime} \widehat{H}_t^{-1} \mathbf{\iota}}.
\end{equation}
We evaluate the following five portfolios, constructed using different covariance matrix estimators:
\begin{enumerate}
    \item  DCC:  the portfolio is defined in \eqref{top}, where \( \widehat{H}_t \) is obtained from a Dynamic Conditional Correlation (DCC) model based on the sample correlation matrix.
    \item  DCC-NL:  the portfolio in \eqref{top}, where \( \widehat{H}_t \) is estimated from the unconditional correlation matrix processed with nonlinear shrinkage, this procedure follows \cite{engle2019large}. 
    \item  DCS (tau): the portfolio in \eqref{top}, where \( \widehat{H}_t \) is estimated from the DCS model in Definition \ref{dcs} using Kendall's tau statistic.
    \item  DCS (rho): the portfolio in \eqref{top}, where \( \widehat{H}_t \) is estimated from the DCS model in Definition \ref{dcs} using Spearman's rho statistic.
    \item  \( 1/p \):  the equal-weighted portfolio, a common benchmark advocated by \cite{demiguel2009optimal}.
\end{enumerate}
For the estimation of the inverse correlation matrix, we start from a rank-based SKEPTIC estimator constructed from Kendall's tau and Spearman's rho statistics. Rank-based dependence measures are substantially more robust than Pearson correlation under heavy tails, skewness, and other departures from Gaussianity, and they provide superior support recovery for the precision matrix in high-dimensional settings \cite{liu2012high}. To obtain a stable and well-conditioned estimate of the inverse correlation matrix, we impose sparsity through the QUIC algorithm of \citet{hsieh2014quic}, selecting the regularization parameter $\lambda$ via the StARS stability criterion of \citet{liu2010stability}\footnote{The tuning parameter is set to $\lambda=0.596$ for the S\&P100, and $\lambda=0.547$ for the S\&P500.}. This approach ensures that only the most reliable conditional dependencies are retained and avoids applying shrinkage to a Pearson-based correlation matrix, which is known to be less reliable in non-Gaussian environments.

We estimate the portfolio weights using the five aforementioned covariance matrix estimators. 
We evaluate the portfolios using the following performance metrics. We include the annualized average value (AV), the annualized standard deviation (SD) of the out-of-sample portfolio returns, and the Sharpe ratio (SR), which is obtained by dividing the portfolio's out-of-sample average value by its standard deviation (SR = AV/SD). 
We also include the turnover (TO) metric to capture the changes in portfolio weights, reflecting the magnitude of portfolio rebalancing costs: 
$$
    \text{TO} = \frac{1}{72}\sum_{h=2}^{73}|w_{h-1} - w_{h}|. 
$$
\begin{table}[!h]
\centering
\renewcommand{\arraystretch}{1.3}
\setlength{\tabcolsep}{6pt}
\begin{tabular}{l|cccc|cccc}
\hline
\hline
\textbf{} & \multicolumn{4}{c|}{\textbf{S\&P100}} & \multicolumn{4}{c}{\textbf{S\&P500}} \\
\hline
\textbf{} & \textbf{AV} & \textbf{SD} & \textbf{SR} & \textbf{TO} & \textbf{AV} & \textbf{SD} & \textbf{SR} & \textbf{TO} \\
\hline
DCC            & 7.17  & 16.57 & 0.43 & 3.86  & 8.49  & 16.38 & 0.52 & 6.96  \\
DCC-NL         & 7.49  & {16.42} & 0.46 & 3.57  & 8.96  & {15.70} & {0.57} & 5.10  \\
DCS (tau)      & 10.00 & 16.45 & 0.61 & 0.47 &  7.68 & 15.81 & 0.48 & 0.83  \\
DCS (rho)      & 10.95 & 17.05 & 0.64 & 0.37 & 9.23 & 17.13 & {0.54} & 0.52  \\
EW             & {14.12} & 20.01 &{0.71} & 0.01  & 11.62 & 21.05 & 0.55 & 0.01  \\
\hline
\hline
\end{tabular}
\caption{Comparison of annualized performance measures (expressed in percentage) for S\&P100 and S\&P500 datasets. AV = Average Value, SD = Standard Deviation, SR = Sharpe ratio, TO = Turnover.}
\label{tab:filtered_performance_metrics}
\end{table}
Table \ref{tab:filtered_performance_metrics} reports the portfolio performance metrics. %The key results reveal that the DCS model using Kendall's tau statistic consistently outperforms the DCC model in terms of average portfolio returns, 
% but exhibits significantly higher volatility. Consequently, the increase in both returns and volatility results in no statistically significant difference in the Sharpe ratio. These findings are further supported by the two-sample means \( t \)-test, the two-tailed \( F \)-test on variance, and the Sharpe ratio test of \citet{ardia2018peer}, with the corresponding results available in the Appendix \ref{tests_ptf}.
The key finding is the higher annualized return achieved by the DCS-based portfolios compared to the DCC-based ones, particularly for portfolios constructed from the S\&P100 constituents. The slightly higher standard deviation of the DCS-based portfolios is compensated by a substantially higher return, resulting in a greater Sharpe ratio.
Additionally, as is predictable, the portfolios constructed using the DCS model show a lower turnover. The ability of the SKEPTIC estimator to recover the true precision matrix in non-Gaussian cases can be exploited for portfolio construction to reduce transaction costs, without sacrificing the portfolio performance. We test portfolio performance metrics in the Appendix \ref{app_test_ptf}.

Nevertheless, for portfolios based on the S\&P500 constituents, the Sharpe ratios of the DCS portfolios are slightly lower than those of the DCC-based and equally weighted portfolios. In contrast, while their standard deviations are not the lowest, they remain sufficiently contained to make the DCS-based portfolios a plausible and stable alternative, further supported by their lower turnover.
\subsection{Portfolio Backtesting}
We evaluate our portfolios in terms of market risk, as measured by the Value at Risk (VaR) and the Expected Shortfall (ES). %We evaluate our portfolio's ability to capture extreme market events, employing Value at Risk (VaR) and Expected Shortfall (ES) as key measures. 
Given a probability level $\alpha$, the VaR estimates the maximum potential daily loss at the specified confidence level $\alpha$, whereas the Expected Shortfall represents the expected loss conditional on the VaR threshold being exceeded.  

Under the normality assumption, given the confidence level $\alpha$, the Value at Risk of the portfolio at $t+1$ is computed as:  
\begin{equation}
\operatorname{VaR}_{t+1}^p=-\sigma_{\operatorname{ptf}, t+1} \cdot \Phi^{-1}(\alpha),
\end{equation}
and the related Expected Shortfall is given by:  
\begin{equation}
\mathrm{ES}=\sigma_{\operatorname{ptf}, t+1} \frac{\varphi\left(\Phi^{-1}(\alpha)\right)}{\alpha},
\end{equation}
where $\sigma_{\operatorname{ptf},t+1}$ denotes the volatility of the portfolio estimated from the out-of-sample return analysis, and $\varphi$ the density of a standard normal distribution. To obtain the dynamic estimate of the portfolio volatility, we fit a GJR-GARCH(1,1) model to portfolio returns. We recall that the portfolio returns are defined as $r_{t+1}^{\mathrm{ptf}}=w_t^{\prime} r_{t+1}$, where $r_{t+1}$ are the realized returns of the stocks that compose the indices, and $r^{ptf}_{t+1}$ are the realized returns of the portfolios.  

We set the confidence level at $\alpha = 5\%$. To assess whether the model and estimation procedure produce reliable risk measure estimates, we conduct three backtesting procedures: the Unconditional Coverage (UC) test (\citet{kupiec1995techniques}), the Conditional Coverage (CC) test (\citet{christoffersen2000relevant}), and the Dynamic Quantile (DQ) test  (\citet{engle2004caviar}). These tests evaluate whether the empirical violation probability aligns with the expected violation probability implied by the confidence level, thereby capturing the realized risk. All three tests are based on the null hypothesis that the expected number of violations matches the observed number.

Table \ref{tableaa} reports the results of the backtesting procedures.
\begin{table}[!h]
    \centering
    \caption{Backtesting procedure results. In the Table, Fail no. indicates the percentage of failures, UC the test statistic values of the unconditional coverage test, CC the test statistic values of the conditional coverage test, and DQ the test statistic values of the dynamic quantile test. \textit{Note:} $^* p<0.1$; $^{**} p<0.05$; $^{***} p<0.01$.}
    \label{tableaa}
    \begin{tabular}{l|cccc|cccc}
    \hline
    \hline
    & \multicolumn{4}{c|}{S\&P 100} & \multicolumn{4}{c}{S\&P 500} \\
    \hline
    \textbf{$VaR^{0.05}_{t+1}$} & Fail no. & UC & CC & DQ & Fail no. & UC & CC & DQ \\
    \midrule
    DCC     & 0.05  & 0.26  & 0.38  & 7.54   & 0.05 & 0.30  & 0.35 & 3.14  \\
    DCC-NL  & 0.05  & 0.15  & 0.94 & 10.31    & 0.05  & 0.30  & 1.02  & 2.85   \\
    DCS Tau & 0.05  & 0.01  & 0.02  & 5.14    & 0.05 & 0.26  & 0.73  & 3.88   \\
    DCS Rho & 0.05  & 0.01  & 0.02  & 4.67    & 0.05  & 0.08  & 0.42  & 3.45  \\
    EW      & 0.05  & 0.54  & 0.60  & 6.25    & 0.05 & 0.26  & 0.95  & 5.30   \\
    \midrule
    \textbf{$ES^{0.05}_{t+1}$} & Fail no. & UC & CC & DQ & Fail no. & UC & CC & DQ \\
    \midrule
    DCC     & 0.03  & 10.19$^{***}$ & 11.07$^{***}$  & 15.40$^{*}$  & 0.03 & 24.99$^{***}$  & 24.99$^{***}$  & 23.88$^{***}$  \\
    DCC-NL  & 0.03  & 11.97$^{***}$ & 12.09$^{***}$  & 17.17$^{**}$ & 0.03  & 19.59$^{***}$ & 19.61$^{***}$  & 20.69$^{***}$ \\
    DCS Tau & 0.03  & 12.89$^{***}$ & 13.09$^{***}$  & 12.61  & 0.03  & 13.92$^{***}$ & 14.13$^{***}$  & 17.85$^{**}$  \\
    DCS Rho & 0.03  & 13.88$^{***}$ & 14.04$^{***}$  & 13.53  & 0.03  & 11.06$^{***}$ & 11.15$^{***}$  & 13.50$^{*}$  \\
    EW      & 0.03  & 18.36$^{***}$ & 20.84$^{***}$  & 17.17$^{**}$ & 0.03  & 19.59$^{***}$ & 21.95$^{***}$  & 20.83$^{***}$ \\
    \hline
    \hline
    \end{tabular}
\end{table}
The models deliver optimal VaR values, with the number of failures being equal to the expected 5\% threshold. For the Expected Shortfall estimates, the values remain consistently below the 5\% benchmark, thereby rejecting the null hypothesis of equality. The test statistic values, in conjunction with the observed number of violations, provide strong evidence of the reliability of our portfolios to effectively capture extreme tail events.
\begin{figure}[p]
    \centering
    \includegraphics[width=\textwidth]{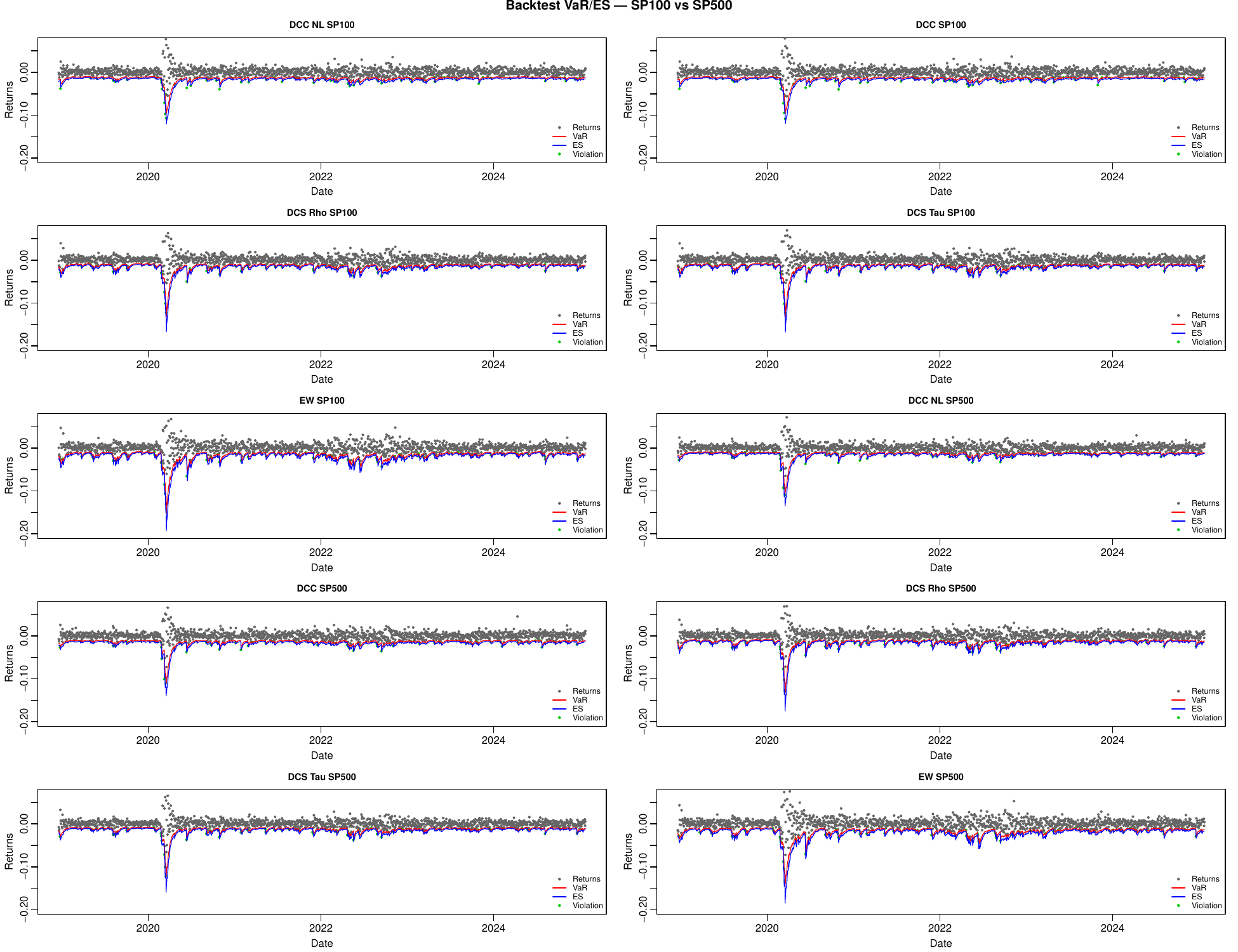}
    \caption{Backtesting results for SP100 and SP500 across the DCC, DCC-NL, DCS (rho), DCS (tau) and EW specifications. Return observations are plotted against their corresponding VaR and ES estimates, with violations highlighted in green.}
    \label{fig:backtest}
\end{figure}

As illustrated in Figure \ref{fig:backtest}, the DCS (tau) and DCS (rho) models exhibit a more flexible and responsive behaviour than both the DCC-based models and the EW benchmark, enabling them to capture shocks in portfolio returns promptly. This flexibility translates into a superior ability to model and predict systemic risk, measured through VaR and Expected Shortfall, with the DCS (tau) model delivering the best overall performance.
\clearpage
\section{Conclusion}\label{section_8}
In this paper, we introduce the Dynamic Conditional SKEPTIC (DCS) model, an extension of the DCC model that relaxes the normality assumption by utilizing nonparametric distributions and rank-based statistics, such as Spearman's rho and Kendall's tau, to estimate the unknown correlation matrix. 

We investigate the theoretical properties of the DCS model, including its stationarity and mixing properties, highlighting its robustness and flexibility in dynamically modeling the correlation matrix. The proposed model provides a more adaptable approach to capture complex market dynamics by addressing the limitations of the normality assumption in asset returns.
%Furthermore, exploring the effects of the nonparametric estimators on model performance could yield valuable insights for enhancing portfolio management strategies.
 A Monte Carlo simulation study demonstrates the effectiveness of the DCS approach in estimating the process parameters.
The empirical analysis demonstrates that the DCS model achieves portfolio Sharpe ratios comparable to benchmark models while having lower turnover, thereby reducing portfolio costs.
In the risk management application, the proposed estimators demonstrate optimal performance with a portfolio that, besides providing optimal average return and low rebalancing costs,  ensures robust market risk tracking.

In summary, the Dynamic Conditional SKEPTIC model represents an appealing alternative to traditional methods for estimating the conditional correlation matrix, offering flexible assumptions and delivering interesting results for portfolios, particularly in the analysis of portfolios based on S\&P100 constituents.
\clearpage

\bibliography{mybibl}
\clearpage
\appendix
\section*{Appendix of Dynamic Conditional SKEPTIC}
\section{Proof of Proposition \ref{proposition_beare}}\label{proof_beare}
We recall the $\left\{W_t\right\}$ process in \eqref{Dcc_mc} that is a stationary Markov chain which implies $\textbf{X}_t$ a stationary process. It is known that its $\beta$-mixing coefficients satisfy:
\begin{equation}
    \beta_k=\frac{1}{2}\left\|F_{0, k}(a, b)-F(a) F(b)\right\|_{\mathrm{TV}},
\end{equation}
where $F_{0, k}$ is the joint distribution function of $X_0$ and $X_k$, and $\|\cdot\|_{\mathrm{TV}}$ is total variation, in the Vitali sense.
We revisit the result of \citet{beare2010copulas}. From Sklar's theorem, and \citet{liu2009nonparanormal} we have
\begin{align*}
    \beta_k&=\frac{1}{2}\|F_{0,k}(a,b)-F(a)F(b)\|_{TV}\\
    & = \frac{1}{2}\|\Phi_{\rho_k}(\Phi^{-1}(a),\Phi^{-1}(b))-\Phi(a)\Phi(b)\|_{TV}\\
    & \le \frac{1}{2}\|\Phi_{\rho_k}(\Phi^{-1}(a),\Phi^{-1}(b))-ab\|_{TV},
\end{align*}
where $\rho_k$ is the correlation between $X_0$ and $X_k$ that depends only to $k$.
Let \( \varphi_k \) denote the density of the semiparametric Gaussian distribution as defined in \eqref{eq:npn_pdf}. This function is differentiable and continuous. It holds that
$
\beta_k \leq \frac{1}{2} \left\| \varphi_k - 1 \right\|_1,
$
and consequently,
$
\beta_k \leq \frac{1}{2} \left\| \varphi_k - 1 \right\|_2.
$
Since \( \varphi_k \) is a symmetric, square-integrable joint density function with uniform marginals, it admits a mean-square convergent expansion:
$$
\varphi_k(a,b)=1+\sum_{i=1}^{\infty} \lambda_i \phi_i(a) \phi_i(b),
$$
where the eigenvalues $\left\{\lambda_i\right\}$ form a nonincreasing square-summable sequence of non-negative real numbers and the eigenfunctions $\left\{\phi_i\right\}$ form a complete orthonormal sequence in $L_2[0,1]$.
We know from \citet{beare2010copulas} that:
\begin{equation}\label{miao}
    C_{k+1}(a,b) = \int_{0}^1 \frac{\partial C_{k}(a,v)}{\partial v} \cdot \frac{\partial C_{1}(v,b)}{\partial v} \, dv,
\end{equation}
which allows us to infer the spectral expansion of the density $\varphi_k$:
\begin{equation}
    \varphi_k(a, b)=1+\sum_{i=1}^{\infty} \lambda_i^k \phi_i(a) \phi_i(b).
\end{equation}
 We now have
\begin{equation}
    \left\|c_k-1\right\|_2=\left\|\sum_{i=1}^{\infty} \lambda_i^k \phi_i(a) \phi_i(b)\right\|_2,
\end{equation}
 so with two applications of Parseval's equality, we obtain
\begin{equation}
   \left\|c_k-1\right\|_2=\left(\sum_{i=1}^{\infty} \lambda_i^{2 k}\right)^{\frac{1}{2}} \leq \lambda_1^{k-1}\left(\sum_{i=1}^{\infty} \lambda_i^2\right)^{\frac{1}{2}}=\lambda_1^{k-1}\|c-1\|_2, 
\end{equation}
where $\lambda_1$ is equal to the maximal correlation of $C$. Since this quantity is assumed to be less than 1, the proof is complete.

\section{Proof of Proposition \ref{dcs_error}}\label{proof_dcs_error}

Fix $m\in\{\tau,\rho\}$ and define
\(
\Delta_t^m
=
\widehat Q_t^m-Q_t^m.
\)
Subtracting the population recursion from its estimated counterpart, we obtain
\[
\Delta_t^m
=
(1-\alpha-\beta)
\left(
\widehat{\bar Q}^{\,m}-\bar Q^m
\right)
+
\beta\Delta_{t-1}^m.
\]

By iterating $\Delta_t^m$, we have
\begin{align*}
\Delta_t^m
&=
\beta^t\Delta_0^m
+
(1-\alpha-\beta)
\sum_{j=0}^{t-1}
\beta^j
\left(
\widehat{\bar Q}^{\,m}-\bar Q^m
\right)\\
&=
\beta^t\Delta_0^m
+
\frac{1-\alpha-\beta}{1-\beta}
(1-\beta^t)
\left(
\widehat{\bar Q}^{\,m}-\bar Q^m
\right).
\end{align*}

The recursions are initialized by plugging their respective long-run
target matrices into the processes. Therefore,
\[
\Delta_0^m
=
\widehat Q_0^m-Q_0^m
=
\widehat{\bar Q}^{\,m}-\bar Q^m.
\]

Hence, $\Delta_t^m$ becomes
\begin{align*}
\Delta_t^m
&=
\beta^t
\left(
\widehat{\bar Q}^{\,m}-\bar Q^m
\right)
+
\frac{1-\alpha-\beta}{1-\beta}
(1-\beta^t)
\left(
\widehat{\bar Q}^{\,m}-\bar Q^m
\right)\\
&=
\left[
\beta^t
+
\frac{1-\alpha-\beta}{1-\beta}
(1-\beta^t)
\right]
\left(
\widehat{\bar Q}^{\,m}-\bar Q^m
\right).
\end{align*}

Therefore, defining
\[
c_t
=
\left[
\beta^t
+
\frac{1-\alpha-\beta}{1-\beta}
(1-\beta^t)
\right],
\]
we have
\[
\left\|
\widehat Q_t^m-Q_t^m
\right\|_{\max}
\leq
c_t
\left\|
\widehat{\bar Q}^{\,m}-\bar Q^m
\right\|_{\max}.
\]

\citet{han2018exponential} derives concentration bounds for rank-based
estimators under temporal dependence. Thus,
\[
\left\|
\widehat{\bar Q}^{\,m}-\bar Q^m
\right\|_{\max}
=
O_p\left(
\sqrt{\frac{\log(Tp)}{T}}
\right).
\]

Since $0\leq\beta<1$, the sequence $c_t$ is uniformly bounded.
Consequently,
\[
\left\|
\widehat Q_t^m-Q_t^m
\right\|_{\max}
=
O_p\left(
\sqrt{\frac{\log(Tp)}{T}}
\right).
\]

Since $R_t^m=f(Q_t^m)$, as defined in
\eqref{nonlinear_transformation}, and, by assumption, the diagonal
elements of $Q_t^m$ and $\widehat Q_t^m$ are uniformly bounded away
from zero and bounded above, the function $f$ is Lipschitz in the
elementwise maximum norm. Therefore, there exists a finite constant
$K$, independent of $t$, such that
\[
\left\|
\widehat R_t^m-R_t^m
\right\|_{\max}
\leq
K
\left\|
\widehat Q_t^m-Q_t^m
\right\|_{\max}.
\]

Consequently,
\[
\left\|
\widehat R_t^m-R_t^m
\right\|_{\max}
=
O_p\left(
\sqrt{\frac{\log(Tp)}{T}}
\right).
\]

Thus, the estimation error of the static SKEPTIC target is propagated
through the stable DCS recursion and the correlation normalization
without changing its stochastic order.

\section{Simulations Results}\label{app_simulation}
\begin{table}[H]
\centering
\caption{Mean and standard deviation of the estimated parameter $\alpha$ and $\beta$ assuming $\delta = 0.01$.}\label{tab_sim_1}
\scriptsize
\renewcommand{\arraystretch}{1.2}
\begin{tabular}{lcccccccc}
\toprule
$\delta = 0.01$ & \multicolumn{2}{c}{DCC} & \multicolumn{2}{c}{DCC NL} & \multicolumn{2}{c}{DCS Rho} & \multicolumn{2}{c}{DCS Tau} \\
\cmidrule(lr){2-3} \cmidrule(lr){4-5} \cmidrule(lr){6-7} \cmidrule(lr){8-9}
$T$ & $\alpha$ & $\beta$ & $\alpha$ & $\beta$ & $\alpha$ & $\beta$ & $\alpha$ & $\beta$ \\
\midrule
\multicolumn{9}{l}{$p = 50$} \\
500  & 0.028 & 0.928 & 0.028 & 0.928 & 0.035 & 0.926 & 0.035 & 0.924 \\
s.d. & 0.004 & 0.013 & 0.004 & 0.013 & 0.005 & 0.006 & 0.005 & 0.006 \\
1000 & 0.025 & 0.941 & 0.026 & 0.942 & 0.029 & 0.939 & 0.029 & 0.938 \\
s.d. & 0.001 & 0.003 & 0.001 & 0.003 & 0.001 & 0.001 & 0.001 & 0.002 \\
2000 & 0.025 & 0.943 & 0.025 & 0.943 & 0.028 & 0.942 & 0.028 & 0.940 \\
s.d. & 0.000 & 0.002 & 0.000 & 0.002 & 0.002 & 0.005 & 0.002 & 0.005 \\
\midrule
\multicolumn{9}{l}{$p = 97$} \\
500  & 0.028 & 0.928 & 0.028 & 0.928 & 0.035 & 0.926 & 0.035 & 0.924 \\
s.d. & 0.004 & 0.013 & 0.004 & 0.013 & 0.005 & 0.006 & 0.005 & 0.006 \\
1000 & 0.025 & 0.941 & 0.026 & 0.942 & 0.029 & 0.939 & 0.029 & 0.938 \\
s.d. & 0.001 & 0.003 & 0.001 & 0.003 & 0.001 & 0.001 & 0.001 & 0.002 \\
2000 & 0.025 & 0.943 & 0.025 & 0.943 & 0.028 & 0.942 & 0.028 & 0.940 \\
s.d. & 0.000 & 0.002 & 0.000 & 0.002 & 0.002 & 0.005 & 0.002 & 0.005 \\
\midrule
\multicolumn{9}{l}{$p = 250$} \\
500  & 0.015 & 0.945 & 0.015 & 0.945 & 0.023 & 0.930 & 0.023 & 0.925 \\
s.d. & 0.004 & 0.010 & 0.004 & 0.010 & 0.002 & 0.008 & 0.002 & 0.008 \\
1000 & 0.016 & 0.947 & 0.016 & 0.947 & 0.022 & 0.941 & 0.022 & 0.939 \\
s.d. & 0.001 & 0.002 & 0.001 & 0.002 & 0.001 & 0.003 & 0.001 & 0.003 \\
2000 & 0.016 & 0.951 & 0.016 & 0.951 & 0.020 & 0.949 & 0.020 & 0.948 \\
s.d. & 0.001 & 0.005 & 0.001 & 0.005 & 0.001 & 0.001 & 0.001 & 0.001 \\
\midrule
\multicolumn{9}{l}{$p = 429$} \\
500  & 0.009 & 0.955 & 0.010 & 0.955 & 0.019 & 0.942 & 0.019 & 0.938 \\
s.d. & 0.002 & 0.003 & 0.002 & 0.003 & 0.002 & 0.001 & 0.002 & 0.001 \\
1000 & 0.014 & 0.944 & 0.014 & 0.944 & 0.018 & 0.945 & 0.018 & 0.943 \\
s.d. & 0.002 & 0.003 & 0.002 & 0.002 & 0.001 & 0.006 & 0.001 & 0.006 \\
2000 & 0.013 & 0.955 & 0.013 & 0.955 & 0.018 & 0.951 & 0.018 & 0.950 \\
s.d. & 0.001 & 0.002 & 0.001 & 0.002 & 0.001 & 0.001 & 0.001 & 0.001 \\
\bottomrule
\end{tabular}
\end{table}
\begin{table}[H]
\centering
\caption{Mean and standard deviation of the estimated parameter $\alpha$ and $\beta$ assuming $\delta = 0.05$.}\label{tab_sim_2}
\scriptsize
\renewcommand{\arraystretch}{1.2}
\begin{tabular}{lcccccccc}
\toprule
$\delta = 0.05$ & \multicolumn{2}{c}{DCC} & \multicolumn{2}{c}{DCC NL} & \multicolumn{2}{c}{DCS Rho} & \multicolumn{2}{c}{DCS Tau} \\
\cmidrule(lr){2-3} \cmidrule(lr){4-5} \cmidrule(lr){6-7} \cmidrule(lr){8-9}
$T$ & $\alpha$ & $\beta$ & $\alpha$ & $\beta$ & $\alpha$ & $\beta$ & $\alpha$ & $\beta$ \\
\midrule
\multicolumn{9}{l}{$p = 50$} \\
500  & 0.024 & 0.936 & 0.024 & 0.937 & 0.031 & 0.929 & 0.031 & 0.926 \\
s.d. & 0.002 & 0.007 & 0.002 & 0.007 & 0.001 & 0.007 & 0.001 & 0.008 \\
1000 & 0.024 & 0.934 & 0.024 & 0.934 & 0.027 & 0.936 & 0.027 & 0.933 \\
s.d. & 0.004 & 0.006 & 0.004 & 0.006 & 0.003 & 0.010 & 0.003 & 0.010 \\
2000 & 0.025 & 0.943 & 0.025 & 0.943 & 0.028 & 0.943 & 0.028 & 0.942 \\
s.d. & 0.001 & 0.001 & 0.001 & 0.001 & 0.001 & 0.001 & 0.001 & 0.001 \\
\midrule
\multicolumn{9}{l}{$p = 97$} \\
500  & 0.021 & 0.930 & 0.022 & 0.930 & 0.026 & 0.935 & 0.027 & 0.930 \\
s.d. & 0.003 & 0.008 & 0.003 & 0.008 & 0.003 & 0.008 & 0.003 & 0.009 \\
1000 & 0.023 & 0.935 & 0.023 & 0.935 & 0.025 & 0.937 & 0.025 & 0.935 \\
s.d. & 0.001 & 0.002 & 0.001 & 0.002 & 0.001 & 0.003 & 0.001 & 0.003 \\
2000 & 0.019 & 0.947 & 0.020 & 0.947 & 0.023 & 0.947 & 0.023 & 0.946 \\
s.d. & 0.001 & 0.003 & 0.001 & 0.003 & 0.001 & 0.005 & 0.001 & 0.005 \\
\midrule
\multicolumn{9}{l}{$p = 250$} \\
500  & 0.014 & 0.941 & 0.014 & 0.942 & 0.020 & 0.937 & 0.020 & 0.932 \\
s.d. & 0.005 & 0.018 & 0.005 & 0.018 & 0.002 & 0.002 & 0.002 & 0.002 \\
1000 & 0.014 & 0.948 & 0.014 & 0.948 & 0.020 & 0.942 & 0.020 & 0.940 \\
s.d. & 0.002 & 0.008 & 0.002 & 0.007 & 0.001 & 0.003 & 0.001 & 0.003 \\
2000 & 0.012 & 0.957 & 0.012 & 0.957 & 0.020 & 0.951 & 0.020 & 0.950 \\
s.d. & 0.005 & 0.009 & 0.005 & 0.009 & 0.002 & 0.004 & 0.002 & 0.004 \\
\midrule
\multicolumn{9}{l}{$p = 429$} \\
500  & 0.009 & 0.950 & 0.010 & 0.950 & 0.020 & 0.931 & 0.020 & 0.926 \\
s.d. & 0.004 & 0.010 & 0.004 & 0.010 & 0.001 & 0.005 & 0.001 & 0.004 \\
1000 & 0.013 & 0.948 & 0.013 & 0.949 & 0.019 & 0.946 & 0.019 & 0.944 \\
s.d. & 0.001 & 0.002 & 0.001 & 0.002 & 0.001 & 0.001 & 0.001 & 0.002 \\
2000 & 0.013 & 0.953 & 0.013 & 0.954 & 0.017 & 0.951 & 0.017 & 0.950 \\
s.d. & 0.001 & 0.001 & 0.001 & 0.002 & 0.001 & 0.001 & 0.001 & 0.001 \\
\bottomrule
\end{tabular}
\end{table}
\begin{table}[H]
\centering
\caption{Mean and standard deviation of the estimated parameter $\alpha$ and $\beta$ assuming $\delta = 0.1$.}\label{tab_sim_3}
\scriptsize
\renewcommand{\arraystretch}{1.2}
\begin{tabular}{lcccccccc}
\toprule
$\delta=0.1$ & \multicolumn{2}{c}{DCC} & \multicolumn{2}{c}{DCC NL} & \multicolumn{2}{c}{DCS Rho} & \multicolumn{2}{c}{DCS Tau} \\
\cmidrule(lr){2-3} \cmidrule(lr){4-5} \cmidrule(lr){6-7} \cmidrule(lr){8-9}
$T$ & $\alpha$ & $\beta$ & $\alpha$ & $\beta$ & $\alpha$ & $\beta$ & $\alpha$ & $\beta$ \\
\midrule
\multicolumn{9}{l}{$p = 50$} \\
500  & 0.025 & 0.930 & 0.026 & 0.931 & 0.032 & 0.923 & 0.032 & 0.917 \\
s.d. & 0.002 & 0.005 & 0.002 & 0.005 & 0.003 & 0.008 & 0.003 & 0.007 \\
1000 & 0.026 & 0.933 & 0.026 & 0.933 & 0.031 & 0.929 & 0.031 & 0.928 \\
s.d. & 0.002 & 0.010 & 0.002 & 0.010 & 0.002 & 0.014 & 0.002 & 0.014 \\
2000 & 0.022 & 0.946 & 0.022 & 0.946 & 0.027 & 0.942 & 0.027 & 0.941 \\
s.d. & 0.003 & 0.007 & 0.003 & 0.007 & 0.002 & 0.003 & 0.002 & 0.002 \\
\midrule
\multicolumn{9}{l}{$p = 97$} \\
500  & 0.021 & 0.931 & 0.021 & 0.931 & 0.029 & 0.923 & 0.029 & 0.919 \\
s.d. & 0.003 & 0.012 & 0.003 & 0.012 & 0.004 & 0.007 & 0.005 & 0.010 \\
1000 & 0.020 & 0.947 & 0.020 & 0.947 & 0.023 & 0.946 & 0.023 & 0.944 \\
s.d. & 0.001 & 0.003 & 0.001 & 0.003 & 0.001 & 0.002 & 0.001 & 0.002 \\
2000 & 0.019 & 0.948 & 0.019 & 0.948 & 0.023 & 0.947 & 0.023 & 0.946 \\
s.d. & 0.002 & 0.004 & 0.002 & 0.004 & 0.001 & 0.003 & 0.001 & 0.003 \\
\midrule
\multicolumn{9}{l}{$p = 250$} \\
500  & 0.012 & 0.947 & 0.012 & 0.947 & 0.021 & 0.933 & 0.021 & 0.929 \\
s.d. & 0.001 & 0.003 & 0.001 & 0.003 & 0.002 & 0.008 & 0.003 & 0.010 \\
1000 & 0.014 & 0.945 & 0.014 & 0.946 & 0.020 & 0.947 & 0.020 & 0.944 \\
s.d. & 0.003 & 0.014 & 0.003 & 0.014 & 0.003 & 0.004 & 0.003 & 0.004 \\
2000 & 0.015 & 0.951 & 0.015 & 0.951 & 0.018 & 0.951 & 0.018 & 0.950 \\
s.d. & 0.001 & 0.003 & 0.001 & 0.003 & 0.000 & 0.001 & 0.000 & 0.001 \\
\midrule
\multicolumn{9}{l}{$p = 429$} \\
500  & 0.012 & 0.947 & 0.012 & 0.947 & 0.021 & 0.933 & 0.021 & 0.929 \\
s.d. & 0.001 & 0.003 & 0.001 & 0.003 & 0.002 & 0.008 & 0.003 & 0.010 \\
1000 & 0.014 & 0.945 & 0.014 & 0.946 & 0.020 & 0.947 & 0.020 & 0.944 \\
s.d. & 0.003 & 0.014 & 0.003 & 0.014 & 0.003 & 0.004 & 0.003 & 0.004 \\
2000 & 0.015 & 0.951 & 0.015 & 0.951 & 0.018 & 0.951 & 0.018 & 0.950 \\
s.d. & 0.001 & 0.003 & 0.001 & 0.003 & 0.000 & 0.001 & 0.000 & 0.001 \\
\bottomrule
\end{tabular}
\end{table}

\clearpage

\section{Tests of Portfolio Performance 
Metrics}\label{app_test_ptf}
In this section, we perform statistical tests to assess the significance of the portfolio metrics.
\subsection{t-test for Average Sample Means}
 Let $\pi_r$ and $\pi_s$ represent two generic portfolios, both with $n$ observations. We perform the two-sample $t$-test. 
Let $r_{t, \pi_r}$ and $r_{t, \pi_s}$ denote the log-return at time $t$ of portfolio $\pi_r$ and $\pi_s$. The two-sample $t$-test is defined as:
$$
\widehat{t}_{\pi_r, \pi_s}=\frac{\Bar{r}_{\pi_r}-\Bar{r}_{\pi_s}}{\sqrt{\frac{s_{\pi_r}^2+s_{\pi_s}^2}{n}}}
$$
where $\Bar{r}_{\pi_r}$ and $\Bar{r}_{\pi_s}$ are, respectively, the average returns of $\pi_r$ and $\pi_s$. 
The values $s_{\pi_r}^2$ and $s_{\pi_s}^2$ are the sample variance of the portfolios, defined as:
\begin{equation}\label{variance}
     s_{\pi, r}^2=\frac{1}{n-1} \sum_{t-1}^n\left(r_{\pi_{r, t}}-\Bar{r}_{\pi_r}\right)^2, \quad s_{\pi, s}^2=\frac{1}{n-1}\sum_{t-1}^n\left(r_{\pi_{s, t}}-\Bar{r}_{\pi_s}\right)^2. 
\end{equation}
In Table \ref{sample_t_test}, we present the results of the $t$ test. 
\begin{table}[!ht]
\centering
\begin{tabular}{lcc}
\hline \hline
 & \multicolumn{1}{c}{{SP100}} & \multicolumn{1}{c}{{SP500}}\\
\textbf{Comparison} & \textbf{$t$-Test Value}  & \textbf{$t$-Test Value}  \\
\hline
  DCC vs DCS (tau) & -0.354 & 0.087\\
  DCC vs DCS (rho) & -0.396 & -0.075\\
  DCC vs DCC NL & -0.034 & -0.052\\
  DCC vs EW & -0.662 & -0.279\\
  DCC NL vs EW & -0.634 & -0.239\\
  DCC NL vs DCS (rho) & -0.364 & -0.026\\
  DCC NL vs DCS (tau) & -0.321 & 0.142\\
  DCS (tau) vs DCS (rho) & -0.043 & -0.162\\
  DCS (tau) vs EW & -0.335 & -0.359\\
  DCS (rho) vs EW & -0.295 & -0.209\\
\hline \hline
\end{tabular}
\caption{Summary of the $t$-tests of annualized average sample means. In the Table, $t$-statistic values are shown. \textit{Note:} $^* p<0.1$; $^{**} p<0.05$; $^{***} p<0.01$.}
\label{sample_t_test}
\end{table}

\subsection{F-test for  Sample Variance}
We recall from \eqref{variance} $s_{\pi_r}^2$, which is the unbiased estimator of variance, and we perform the $F$-test on the ratio of variances:
$$
F=\frac{s_{\pi_r}^2}{s_{\pi_s}^2}, \quad H_0: F=1
$$
In Table \ref{sample_F_variance}, the results of the $F$-test for the variances of the portfolios are shown. 
\begin{table}[!ht]
\centering
\begin{tabular}{lcc}
\hline \hline
 & \multicolumn{1}{c}{{SP100}} & \multicolumn{1}{c}{{SP500}} \\
\textbf{Comparison} & \textbf{$F$-Test Value} & \textbf{$F$-Test Value} \\
\hline
  DCC vs DCS (tau) & 0.947 & 1.068\\
  DCC vs DCS (rho) & 0.942 & 0.924\\
  DCC vs DCC NL & 1.018 & $1.089^{*}$\\
  DCC vs EW & $0.686^{***}$ & $0.62^{***}$\\
  DCC NL vs EW & $0.674^{***}$ & $0.57^{***}$\\
  DCC NL vs DCS (rho) & 0.926 & $0.849^{***}$\\
  DCC NL vs DCS (tau) & 0.93 & 0.981\\
  DCS (tau) vs DCS (rho) & 0.995 & $0.865^{***}$\\
  DCS (tau) vs EW & $0.724^{***}$ & $0.581^{***}$\\
  DCS (rho) vs EW & $0.728^{***}$ & $0.671^{***}$\\
\hline 
\end{tabular}
\caption{Summary of the $F$-tests of annualized rate of variances. In the Table, $F$-statistic values are shown. \textit{Note:} $^* p<0.1$; $^{**} p<0.05$; $^{***} p<0.01$.}
\label{sample_F_variance}
\end{table}

\subsection{Sharpe ratio Test}
We perform the Sharpe ratio test using the "PeerPerformance" \texttt{R} package described in \citet{ardia2018peer}, with a block bootstrap approach.
Table \ref{sample_SR} presents the results of the Sharpe ratio test. 
\begin{table}[h!]
\centering
\begin{tabular}{lcc}
\hline \hline
& \multicolumn{1}{c}{SP100} & \multicolumn{1}{c}{SP500} \\
\textbf{Comparison} & \textbf{$\Delta$ Sharpe ratio} & \textbf{$\Delta$ Sharpe ratio} \\
\hline
DCC vs DCS (tau)         & $-0.012$ & $0.002$ \\
DCC vs DCS (rho)         & $-0.013$ & $-0.001$ \\
DCC vs DCC NL            & $0.001$  & $0.003$ \\
DCC vs EW                & $-0.017$ & $-0.002$ \\
DCC NL vs EW             & $-0.016$ & $0.001$ \\
DCC NL vs DCS (rho)      & $-0.012$ & $0.002$ \\
DCC NL vs DCS (tau)      & $-0.010$ & $0.005$ \\
DCS (tau) vs DCS (rho)   & $0.001$  & $0.004$ \\
DCS (tau) vs EW          & $-0.006$ & $-0.004$ \\
DCS (rho) vs EW          & $-0.004$ & $-0.001$ \\
\hline \hline
\end{tabular}
\caption{Sharpe ratio test following \citet{ardia2018peer}. The table reports $\Delta$ Sharpe ratios (annualized), i.e., the difference between the two portfolios' Sharpe ratios in each comparison. \textit{Note:} $^* p<0.1$; $^{**} p<0.05$; $^{***} p<0.01$.}
\label{sample_SR}
\end{table}

\clearpage

\end{document}